\documentclass[sort&compress]{article}
\usepackage{graphicx}
\usepackage{bm}
\usepackage{citesort}
\usepackage{amsmath}
\usepackage{amssymb}
\makeatletter
\title{Guiding structures with multiply connected cross-sections: evolution of propagation in external fields at complex Robin parameters}
\author{O. Olendski \\
King Abdullah Institute for Nanotechnology\\
King Saud University\\
P.O. Box 2454, Riyadh 11451, Saudi Arabia\\
oolendski@ksu.edu.sa\\
}
\begin{document} 

\maketitle
\begin{abstract}
Properties of the two-dimensional ring and three-dimensional infinitely long straight hollow waveguide with unit width and inner radius $\rho_0$ in the superposition of the longitudinal uniform magnetic field $\bf B$ and  Aharonov-Bohm flux are analyzed within the framework of the scalar Helmholtz equation under the assumption that the Robin boundary conditions at the inner and outer confining walls contain extrapolation lengths $\Lambda_{in}$ and $\Lambda_{out}$, respectively, with nonzero imaginary parts. It is shown that, compared to the disk geometry, the annulus opens up additional possibilities of varying magnetization and currents by tuning imaginary components of the Robin parameters on each confining circumference; in particular, the possibility of restoring a lossless  longitudinal flux by zeroing imaginary part $E_i$ of the total transverse energy $E$ is discussed. The energy $E$ turns real under special correlation between the imaginary parts of $\Lambda_{in}$ and $\Lambda_{out}$ with the opposite signs what corresponds to the equal transverse fluxes through the inner and outer interfaces of the annulus. In the asymptotic case of the very large radius, simple expressions are derived and applied to the analysis of the dependence of the real energy $E$ on $\Lambda_{in}$ and $\Lambda_{out}$. New features also emerge in the magnetic field influence; for example, if, for the quantum disk, the imaginary energy $E_i$ is quenched by the strong intensities $B$, then for the annulus this takes place only when the inner Robin distance $\Lambda_{in}$ is real; otherwise, it almost quadratically depends on  $B$ with the corresponding enhancement of the reactive scattering. Closely related problem of the hole in the otherwise uniform medium is also addressed for real and complex extrapolation lengths with the emphasis on the comparative analysis with its dot counterpart.
\end{abstract}

\section{Introduction}
\label{sec1}
Recent analysis \cite{Olendski1} extended into the whole {\em complex} plane earlier theoretical research \cite{Balian1,Balian2,Balian3} on the influence of the {\em real} Robin parameter $\Lambda$ \cite{Gustafson1} on the properties of spatially confined two- (2D) and three-dimensional (3D) domain $\it\Omega$. Obtained dependencies on the example of the infinitely long cylinder and its circular cross-section generalized previously known results for different {\em real} $\Lambda$ without \cite{Constantinou2,Bosma1,Sieber1,Romeo1,Berry2,Olendski2} or with \cite{Dingle1,Dingle2,SaintJames1,Dalmasso1,Takacs1,Takacs2,Nakamura1,Buisson1,Geerinckx1,Constantinou1,Constantinou2,Avishai1,Benoist1,Ludu1} axial magnetic field $\bf B$. Robin parameter also known as the extrapolation length appears in the boundary condition at the sample interface $\it\partial\Omega$ with its inward normal unit vector $\bf n$ 
\begin{equation}\label{Robin1}
\left.{\bf n}{\bm\nabla}\Psi\right|_{\it\partial\Omega}=\left.\frac{1}{\Lambda}\Psi\right|_{\it\partial\Omega}
\end{equation}
for the function $\Psi(\bf r)$ being a solution of the scalar Helmholtz wave equation
\begin{equation}\label{WaveEquation1}
{\bm\nabla}^2\Psi({\bf r})+k^2\Psi({\bf r})=0.
\end{equation}
Here, the wave vector $k$ is equal to the ratio of the frequency of the electromagnetic or acoustical oscillations $\omega$ to the corresponding speed of propagation $c$
\begin{equation}\label{WaveVector2}
k=\frac{\omega}{c}
\end{equation}
while for the quantum mechanical particle with mass $m_p$ it is expressed as
\begin{equation}\label{WaveVector1}
k=\sqrt{2m_pE}/\hbar,
\end{equation}
where $E$ is a total energy of the particle and $\hbar$ is the reduced Planck constant. In the former case, a solution $\Psi(\bf r)$ is  an acoustical or electromagnetic potential through which an air pressure and velocity or intensities of the electric and magnetic fields are expressed while in the latter case it is a wave function with its square defining the probability of the location of the particle and being normalized according to
\begin{equation}\label{normalization1}
\int_\Omega\left|\Psi\left({\bf r}\right)\right|^2d{\bf r}=1.
\end{equation}
The same boundary condition (\ref{Robin1}) applies also \cite{Zaitsev2,Andryushin1,Andryushin2,Lu1,Lykov1} to the solution of the nonlinear Ginzburg-Landau (GL) equation \cite{Ginzburg1}
\begin{equation}\label{NonlinearEquation1}
{\bm\nabla}^2\Psi({\bf r})+k^2\Psi({\bf r})+\beta\left|\Psi({\bf r})\right|^2\Psi({\bf r})=0
\end{equation}
with positive GL parameter $\beta$. Despite its venerable age, phenomenological GL theory of superconductivity \cite{Ginzburg1,deGennes1} continues to be a powerful tool of studying superconductors as its predictions are in a very good agreement with the experiment. Order parameter $\Psi(\bf r)$ from (\ref{NonlinearEquation1}) defines the density of superconducting particles $n_s$,
\begin{equation}\label{density1}
n_s=|\Psi({\bf r})|^2,
\end{equation}
while the role of the energy $E$ is played by the GL parameter $-\alpha$ which is expressed via the actual temperature $T$ of the superconducting material, the bulk critical temperature at zero magnetic field $T_c$, zero-temperature coherence length $\xi\left(0\right)$ and Cooper pair mass being equal to the double bare electron mass $m_e$:
\begin{equation}\label{Energy1}
E\equiv -\alpha=\frac{\hbar^2}{2m_p\xi^2\left(0\right)}\left(1-\frac{T}{T_c}\right).
\end{equation}
Thus, minimizing the lowest eigenenergy leads to the increase of the critical temperature $T$.

It is worthwhile to note that de Gennes distance, as the extrapolation length $\Lambda$ is called in the language of this field of scientific research, can take negative values for the border with the other superconductor with higher critical temperature what leads to the increase of its own $T$. This theoretically predicted enhancement of superconductivity \cite{Fink1,Montevecchi1,Montevecchi2,Olendski2} was indeed observed in cold worked In$_{0.993}$Bi$_{0.007}$ foils \cite{Fink1} and tin samples \cite{Kozhevnikov1,Kozhevnikov2}.

If a magnetic field $\bf B$ is applied to the system, its influence is included into Eqs.\ (\ref{WaveEquation1}) and (\ref{NonlinearEquation1}) via the vector potential $\bf A$:
\begin{equation}\label{Magnetic1}
\left({\bm\nabla}-i\frac{q}{\hbar}{\bf A}\right)^2\Psi({\bf r})+k^2\Psi({\bf r})=0,
\end{equation}
\begin{equation}\label{NonlinearEquation2}
\left({\bm\nabla}-i\frac{q}{\hbar}{\bf A}\right)^2\Psi({\bf r})+k^2\Psi({\bf r})+\beta\left|\Psi({\bf r})\right|^2\Psi({\bf r})=0,
\end{equation}
where $q$ is the particle charge; in particular, for the Cooper pair, $q=-2e$, with $e$ being the absolute value of the electronic charge. Here, 
\begin{equation}\label{VectorPotential1}
{\bf B}={\bm\nabla}\times{\bf A}.
\end{equation}
Vector potential enters also the GL boundary condition:

\begin{equation}\label{Robin2}
\left.{\bf n}\left({\bm\nabla}-i\frac{q}{\hbar}{\bf A}\right)\Psi\right|_{\it\partial\Omega}=\left.\frac{1}{\Lambda}\Psi\right|_{\it\partial\Omega}.
\end{equation}
In the GL theory \cite{Ginzburg1,deGennes1}, Eq.\ (\ref{NonlinearEquation2}) with boundary condition (\ref{Robin2}) has to be solved together with an expression for the supercurrent density ${\bf j}_s$
\begin{equation}\label{SuperCurrentDensity1}
{\bf j}_s=i\frac{q\hbar}{2m_p}\left(\Psi{\bm\nabla}\Psi^\ast-\Psi^\ast{\bm\nabla}\Psi\right)-\frac{q^2}{m_p}{\bf A}\left|\Psi\right|^2
\end{equation}
and the Maxwell equation
\begin{equation}\label{Maxwell1}
{\bm\nabla}\times{\bf B}=\mu{\bf j}_s,
\end{equation}
$\mu$ is a magnetic permeability. System of equations (\ref{NonlinearEquation2}), (\ref{SuperCurrentDensity1}), (\ref{Maxwell1}) was analyzed in the Neumann limit, $\Lambda=\infty$, of condition (\ref{Robin2}) for the solid cylinder  \cite{Fink3,Fink4,delaCruz1,Moshchalkov2,Zharkov1,Zharkov2,Zharkov3,Zharkov4} and for studied below geometry of its hollow counterpart \cite{Douglas1,Fink2} while the calculations of the combined influence of the magnetic field and real de Gennes distance based on either linear, Eq. (\ref{Magnetic1}), \cite{Hurault1,Takacs1,Takacs2,Buisson1,Masale1,Hornberger1,Hornberger2,Slachmuylders1,Zhu1,Zha1,Calero1,Moncada1,Ludu1} or nonlinear, Eq. (\ref{NonlinearEquation2}), \cite{Richardson1,Baelus2,Pogosov1,Pan1,Kachmar1,Kachmar2,Kachmar3,Kachmar4,Kachmar5} GL theory revealed, for different shapes, a significant influence of the parameter $\Lambda$ on the nucleation of the superconductivity, critical magnetic fields and localization properties of the superconducting state. We remark that the linearised GL equation (\ref{Magnetic1}) which is the main subject of the present study correctly captures the physical phenomena in the uniform magnetic field $\bf B$ and temperature $T$ ranges close to the transition to the normal state when the order parameter $\Psi({\bf r})$ is small and, accordingly, the cubic term in (\ref{NonlinearEquation2}) can be safely neglected  \cite{deGennes1}.  The influence of the cubic term can be strongly suppressed also by the choice of the metal or alloy \cite{Olendski3} since the GL parameter $\beta$ contains material-dependent density of states at the Fermi energy $N(0)$, coherence length $\xi(0)$, critical temperature $T_c$ and the mean free path $l$ \cite{deGennes1}. In addition, it can be shown that equations derived below for the linear case follow also from the complete nonlinear GL theory \cite{Zhu1,Zha1}. Moreover, a comparison between the theory \cite{Deo1} and experiment \cite{Geim1} revealed that, in the Neumann limit, $\Lambda=\infty$, the linearized GL equation correctly captures features of the aluminum disks of the different radii when the predictions of the full GL theory sometimes are in a worse agreement with the experiment \cite{Geim1} than the results produced by the solution of Eq. (\ref{Magnetic1}). Thus, the analysis of the linear magnetic Helmholtz equation is indispensable in the study of the properties of superconductors.

If the extrapolation length $\Lambda$ can take negative values, a natural question arises: what happens if the Robin parameter is {\em complex}? Attempts to answer it have been made during the study of the scattering phenomena in different media \cite{Lakshtanov1} such as a sound duct with porous lining \cite{Ko1,Ko2,Rostafinski1,Rienstra1,Rienstra2,Felix1,Bi1,Allard1}, impedance electromagnetic waveguides \cite{Katsenelenbaum1}, ferrite-filled resonator systems \cite{Krupka1}, absorbers in high-frequency electromagnetic scattering \cite{Weston1}. A comprehensive answer showed that for the infinitely long cylinder with singly connected circular cross section the imaginary part of the transverse {\em complex} energy $E^\perp$ exhibits a pronounced maximum as a function of the imaginary part $\Lambda_i$ of the de Gennes distance \cite{Olendski1}. The energy  $E^\perp$ is a transverse component of the total energy $E$ entering equation (\ref{WaveEquation1}) via (\ref{WaveVector1}). As a result, the current undergoes a resonant alteration as it  flows down the wire exponentially increasing or decreasing (depending on the sign of $\Lambda_i$) with the longitudinal distance. This change of the longitudinal flux is accompanied by the transverse radial currents through the circumference of the disk absent for the real de Gennes distances \cite{deGennes1}. Nonzero real part $\Lambda_r$ of the Robin parameter and axial uniform magnetic field $\bf B$ quench the resonance and, for their large values, restore the lossless longitudinal current. Physically, it is explained by the fact that, for example,  the increasing magnetic field squeezes the charged particle to the cylinder axis; accordingly, with magnetic intensity growing the influence of the boundary decreases and transverse energies $E^\perp$ transform, for the large $B$, into the Landau levels.

Apart from mentioned above acoustical and electromagnetic systems, complex extrapolation lengths can be realized for the superconducting materials too \cite{Olendski1}. Namely, it was argued recently \cite{Lipavsky1,Morawetz1,Kolacek1} that the electric field $\mbox{\boldmath$\cal E$}$ applied perpendicularly to the surface should be accounted for in the total de Gennes distance $\Lambda_{tot}$ by the addition to the inverse zero-field extrapolation length $1/\Lambda$ of the extra term ${\cal E}/U_s$
\begin{equation}\label{TotalExtrapolationLength1}
\frac{1}{\Lambda_{tot}}=\left.\frac{1}{\Lambda}\right|_{{\cal E}=0}+\frac{\cal E}{U_s}
\end{equation}
with the potential $U_s$ being expressed through the parameters of the GL theory:
\begin{equation}\label{PotentialU}
\frac{1}{U_s}\cong\kappa^2\frac{\partial\ln T_c}{\partial\ln n_s}\frac{\left|q\right|\epsilon_s}{m_pc^2}.
\end{equation}
Here, dimensionless GL parameter $\kappa$ is the ratio of the zero-temperature London penetration length $\lambda(0)$ to the coherence length, $\kappa=\lambda(0)/\xi(0)$; $\epsilon_s$ is superconductor ionic background permittivity, and $c$ is speed of light. If the permittivity $\epsilon_s$ has a noticeable imaginary part, so does, according to (\ref{TotalExtrapolationLength1}) and (\ref{PotentialU}), the total de Gennes distance too. In this model, a variation of the imaginary part of the Robin parameter is achieved by a simple change of an applied gate voltage.

Considered in Ref.\ \cite{Olendski1} model of the solid cylinder offers only one surface through which the transverse flux can enter or leave the sample. One can expect that the multiply connected structure with at least two confining interfaces offers additional channel(s) of controlling its transport and thermodynamical properties. The main subject of the present research is to investigate the interaction of the inner and outer complex Robin parameters and its influence on the conductivity and magnetization of the hollow cylindrical waveguide and its 2D annular cross section. It is shown that, indeed, varying the signs and magnitudes of the de Gennes distances on each confining wall, one can manipulate these properties in a wide range; in particular, the conditions for the lossless longitudinal flow down the channel are derived and analyzed:  if the incoming transverse torrent through one surface with complex $\Lambda$ is equal in magnitude to the outgoing flow through the other wall with the opposite sign of the imaginary part of its extrapolation length, then the longitudinal current does not change along the duct. New features emerge also when the magnetic field is applied parallel to the channel axis; namely, depending on the sign and magnitude of the inner de Gennes distance, the alteration of the current can be manipulated at will by the appropriate change of $\bf B$. For example, the lossless current down the waveguide at the strong magnetic intensities is asymptotically achieved only for the real inner Robin parameter; otherwise, since the field pushes the charged carrier closer to the reactively scattering inner surface, the attenuation or amplification of the longitudinal flux almost quadratically depends on the field. As a by-product, not considered before aspects of the problem of the hole in the film are discussed too and its comparative analysis with the solid cylinder case is performed.

The paper is organized as follows. In Section II our model is presented and a necessary formulation of our method is given. Section III is devoted to the presentation and detailed mathematical and physical interpretation of the calculated results. Summary of the research is provided in Section IV.
\begin{figure}
\centering
\includegraphics[width=0.7\columnwidth]{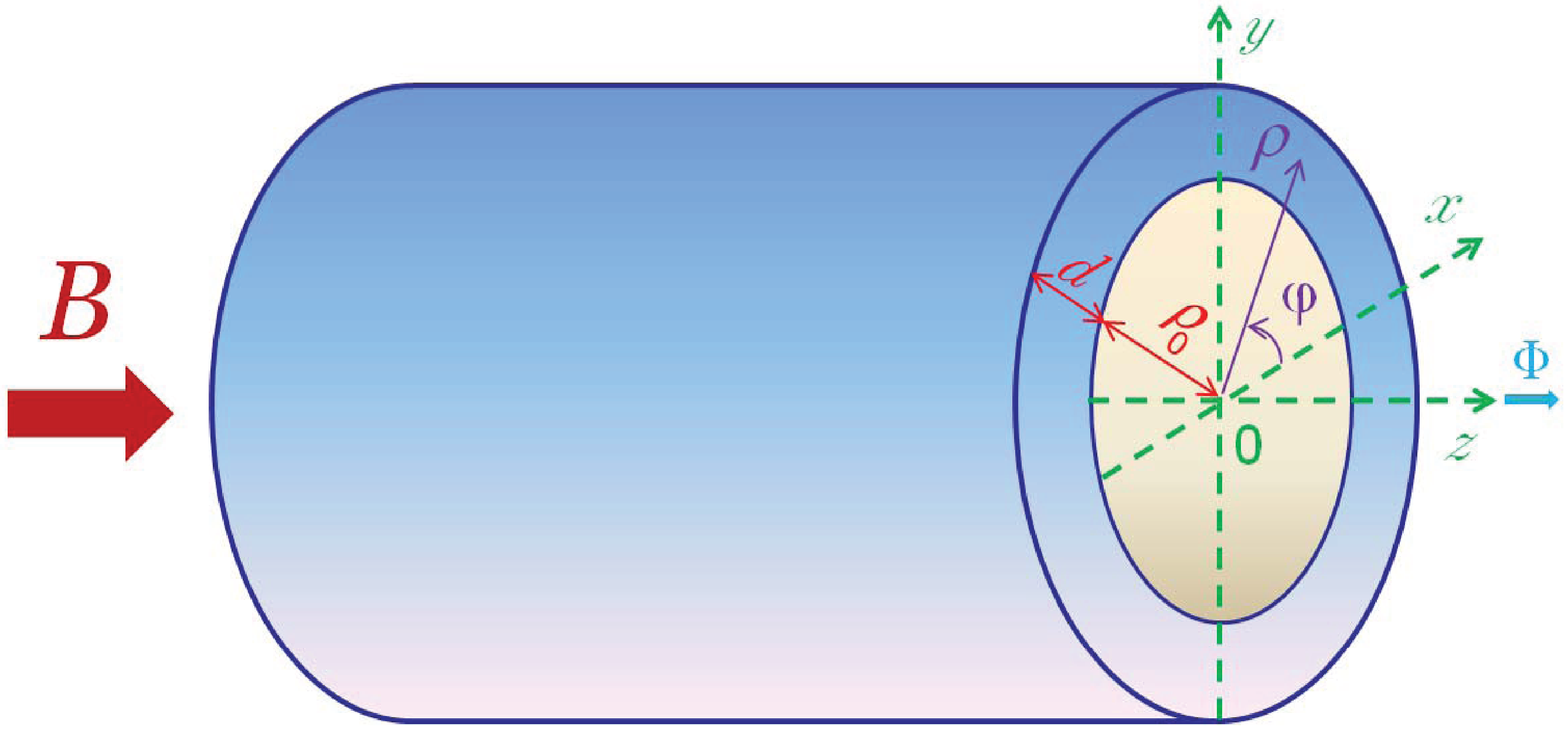}
\caption{\label{Fig1}
Schematic of the infinitely long hollow superconductor wire of the inner radius $\rho_0$ and the outer one $(\rho_0+d)$ subjected to the superposition of the uniform magnetic field $\bf B$ and the AB flux $\bf\Phi$ parallel to its axis. Channel walls support Robin boundary condition, Eq.~(\ref{Robin1}), with the complex outer $\Lambda_{out}$ and inner $\Lambda_{in}$ extrapolation lengths uniform along the corresponding wire surface. Cartesian $(x,y,z)$ and cylindrical $(\rho,\varphi,z)$ systems of coordinates are also shown with their origins lying on the waveguide axis. Curved arrow shows the azimuthal direction in which the polar angle $\varphi$ grows.
}
\end{figure}
\section{Model and formulation}
\label{sec2}
Infinitely long 3D straight wire of the annular cross section with the inner $\rho_0$ and outer $(\rho_0+d)$ radii is placed into the uniform magnetic field $\bf B$ with its direction coinciding with the channel axis (Fig. \ref{Fig1}). For the zero inner length, $\rho_0=0$, one recovers the solid cylinder of the radius $d$ treated before \cite{Olendski1} (note different length notations here and in Ref. \cite{Olendski1}).  Another asymptotics of the infinite annulus width, $d=\infty$, transforms the system into the columnar defect of the radius $\rho_0$ in the otherwise uniform  medium that will be addressed below too. Cylindrical walls of the hollow waveguide support boundary condition, as described by Eq.~(\ref{Robin1}) with the uniform along the length and circumference extrapolation lengths $\Lambda_{in}$ for the inner surface and $\Lambda_{out}$ for the outer one. We do not confine the values of $\Lambda_{in,out}$ to be real concentrating on the properties of the structure at the complex de Gennes distances. For completeness, we also introduce the Aharonov-Bohm (AB) whisker \cite{Aharonov1,Skarzhinskiy1,Afanasev1} with its total magnetic flux $\bf\Phi$ coinciding with the cylinder axis. Similar to the solid cylinder case \cite{Olendski1}, our analysis will be based on linear Helmholtz equation (\ref{Magnetic1}) what, in the case of superconductors, together with the fact that the background magnetic field $\bf B$ is uniform means that we restrict our consideration to the range of the magnetic intensities and temperatures close to the transition between superconducting and normal states \cite{deGennes1} even though the results obtained are covered by the full GL theory (see Ref.~\cite{Olendski1} for more discussion) and have much wider validity range.

Geometry of the system dictates a natural choice of the cylindrical system of coordinates ${\bf r}\equiv(\rho,\varphi,z)$ with its origin lying at the circles center and the $z$ axis being parallel  to the waveguide. We will seek the solutions of Eq.~(\ref{Magnetic1}) with the vector potential written in the symmetrical gauge, 
\begin{equation}\label{VectorPotential3}
{\bf A}=\frac{1}{2}\left[\left({\bf B}+\frac{{\bm\Phi}}{\pi\rho^2}\right)\times{\bf r}\right].
\end{equation}
Then, one has
\begin{equation}\label{VectorPotential2}
{\bf A}=\left(0,\frac{1}{2}B\rho+\frac{\Phi}{2\pi\rho},0\right).
\end{equation} 

We will operate with the energy $E$ through which the wave vector $k$ is expressed, according to (\ref{WaveVector1}). Such a treatment describes a superconductor wire. A transition to the frequency $\omega$ of the acoustical or electromagnetic oscillations can be readily done with the help of (\ref{WaveVector2}). It is convenient at this point to choose dimensionless variables; namely, we will measure all lengths in units of the width of the ring $d$; accordingly, if not stated otherwise, all energies will be measured in units of the ground-state energy $\pi^2\hbar^2/(2m_pd^2)$ of the infinite Dirichlet 1D quantum well of width $d$; all momenta, in units $1/d$; magnetic fields, in units of $\hbar/(|q|d^2)$; magnetization, in units of $\hbar|q|/(2m_p)$; 2D current density, in units of $q\hbar/(m_pd^4)$; current, in units of $q\hbar/(m_pd)$; time, in units of $2m_pd^2/(\pi^2\hbar)$; magnetic flux, in units $h/|q|$. Discussion on the choice of the units and its relation to the description of the processes in different physical systems can be found in Ref. \cite{Olendski1}. Then, Eq.~(\ref{WaveVector1}) in the chosen units transforms to $k=\pi\sqrt{E}$, and Eq.~(\ref{Magnetic1}) for the order parameter $\Psi(\Phi;\rho,\varphi,z)$ becomes:
\begin{equation}\label{WaveEquation2}
\frac{1}{\rho}\frac{\partial}{\partial\rho}\left(\rho\frac{\partial\Psi}{\partial\rho}\right)+\frac{1}{\rho^2}\frac{\partial^2\Psi}{\partial\varphi^2}+\frac{2i}{\rho}\left(\frac{1}{2}B\rho+\frac{\Phi}{\rho}\right)\frac{\partial\Psi}{\partial\varphi}-\left(\frac{1}{2}B\rho+\frac{\Phi}{\rho}\right)^2\Psi+\frac{\partial^2\Psi}{\partial z^2}+\pi^2E\Psi=0
\end{equation}
with $E$ being a total energy of the particle. Factoring out the $z$-dependence
\begin{equation}\label{Factoring1}
\Psi(\Phi;\rho,\varphi,z)=e^{ik_zz}\psi(\Phi;\rho,\varphi)
\end{equation}
leads to the equation for the transverse function $\psi(\Phi;\rho,\varphi)$:
\begin{equation}\label{WaveEquation3}
\frac{1}{\rho}\frac{\partial}{\partial\rho}\left(\rho\frac{\partial\psi}{\partial\rho}\right)+\frac{1}{\rho^2}\frac{\partial^2\psi}{\partial\varphi^2}+\frac{2i}{\rho}\left(\frac{1}{2}B\rho+\frac{\Phi}{\rho}\right)\frac{\partial\psi}{\partial\varphi}-\left(\frac{1}{2}B\rho+\frac{\Phi}{\rho}\right)^2\psi+\pi^2E^\perp\psi=0.
\end{equation}
Longitudinal wave vector $k_z$ and the transverse energy $E^\perp$ are related as
\begin{equation}\label{relation1}
k_z=\pi\sqrt{E-E^\perp}.
\end{equation}
Rotational symmetry of the system allows one to separate out the angular and radial dependencies in the transverse function $\psi(\Phi;\rho,\varphi)$:
\begin{equation}\label{Factoring2}
\psi_{nm}(\Phi;\rho,\varphi)=\frac{1}{\sqrt{2\pi}}e^{im\varphi}R_{nm}(\Phi;\rho).
\end{equation}
Here, $m=0,\pm 1,\pm 2,\ldots$ and $n=0,1,\ldots$ are the azimuthal and the principal quantum numbers, respectively. In this way one arrives at the equation for the radial function $R_{nm}(\Phi;\rho)$:
\begin{equation}\label{RadialEquation1}
\frac{d^2R_{nm}}{d\rho^2}+\frac{1}{\rho}\frac{dR_{nm}}{d\rho}-\left(\frac{m+\Phi}{\rho}+\frac{1}{2}B\rho\right)^2R_{nm}+\pi^2E_{\Phi;nm}^\perp R_{nm}=0.
\end{equation}
This equation is supplemented by the boundary conditions for the function $R_{nm}(\Phi;\rho)$. Our choice of the vector potential in the form of (\ref{VectorPotential2}) drops it out from (\ref{Robin2}) which becomes:
\begin{equation}\label{Robin3}
\left.\left(\frac{dR_{nm}}{d\rho}\mp\frac{1}{\Lambda_{\mbox{\hspace{-2mm}}\,\tiny\begin{array}{l}in\\  out\end{array}}}R_{nm}\right)\right|_{\tiny\begin{array}{l}\rho=\rho_0\\ \rho=\rho_1\end{array}}=0.
\end{equation}

Eqs.\ (\ref{RadialEquation1}) and (\ref{Robin3}) constitute the problem of finding the eigenfunctions $R_{nm}(\Phi;\rho)$ and eigenenergies $E^\perp_{\Phi;nm}$ of the 2D circular annulus with its circumferences supporting the boundary conditions with, in general, complex $\Lambda_{in,out}$.

Analytical solution to Eq.~(\ref{RadialEquation1}) is expressed via the Kummer confluent hypergeometric functions $M(a,b,x)$ and $U(a,b,x)$ \cite{Abramowitz1,Bateman1,Buchholz1}:
\begin{eqnarray}
R_{nm}(\Phi;\rho)=\gamma_{\Phi;nm}\exp\left(-\frac{B}{4}\rho^2\right)\left(\frac{B}{2}\rho^2\right)^{|m_\Phi |/2}\nonumber\\
\label{RadialFunction1}
\left[f_U\left(\rho_1,\Lambda_{out}\right)M\left(a_{m_\Phi},|m_\Phi |+1,\frac{B}{2}\rho^2\right)-f_M\left(\rho_1,\Lambda_{out}\right)U\left(a_{m_\Phi},|m_\Phi |+1,\frac{B}{2}\rho^2\right)\right]
\end{eqnarray}
with
\begin{eqnarray}
\rho_1&=&\rho_0+1,\\
m_\Phi&=&m+\Phi,\\
\label{A1}
a_{m_\Phi}&=&\frac{m_\Phi +|m_\Phi |+1}{2}-\frac{\pi^2}{2}\frac{E_{\Phi;nm}^\perp}{B},
\end{eqnarray}
and functions $f_M(\rho,\Lambda)$ and $f_U(\rho,\Lambda)$ written as
\begin{subequations}\label{FunctionsF}
\begin{eqnarray}\label{FunctionFu}
f_M\left(\rho,\Lambda\right)=\left(\frac{|m_\Phi |}{\rho}-\frac{B}{2}\rho+\frac{1}{\Lambda}\right)M\left(a_{m_\Phi},|m_\Phi |+1,\frac{B}{2}\rho^2\right)+B\rho M'\left(a_{m_\Phi},|m_\Phi |+1,\frac{B}{2}\rho^2\right),\quad\\
\label{FunctionFm}
f_U\left(\rho,\Lambda\right)=\left(\frac{|m_\Phi |}{\rho}-\frac{B}{2}\rho+\frac{1}{\Lambda}\right)U\left(a_{m_\Phi},|m_\Phi |+1,\frac{B}{2}\rho^2\right)+B\rho U'\left(a_{m_\Phi},|m_\Phi |+1,\frac{B}{2}\rho^2\right).\quad
\end{eqnarray}
\end{subequations}
Here, prime denotes a derivative of the function with respect to the last argument and the real coefficient $\gamma_{\Phi;nm}$ is determined from the normalization condition which is either of the form from (\ref{density1}) for superconductors or  
\begin{equation}\label{normalization2}
\int_{\rho_0}^{\rho_0+1}\left|R_{\Phi;nm}(\rho)\right|^2\rho d\rho=1
\end{equation}
for the electromagnetic and acoustical waveguides.

The form of the function $R_{nm}(\Phi;\rho)$ from Eq.~(\ref{RadialFunction1}) automatically satisfies condition (\ref{Robin3}) at the outer surface, $\rho=\rho_0+1$. Imposing boundary requirement at the inner edge, $\rho=\rho_0$, one arrives at the following transcendental equation for determination of the energies $E_{\Phi;nm}^\perp$:
\begin{equation}\label{MagneticEq1}
f_M\left(\rho_0,-\Lambda_{in}\right)f_U\left(\rho_1,\Lambda_{out}\right)-f_U\left(\rho_0,-\Lambda_{in}\right)f_M\left(\rho_1,\Lambda_{out}\right)=0.
\end{equation}

In the limit of the vanishing background magnetic field, $B\rightarrow 0$, Eqs. (\ref{RadialFunction1}), (\ref{FunctionsF}) and  (\ref{MagneticEq1}) transform to (\ref{RadialFunction2}), (\ref{FunctionsB}) and (\ref{BesselEq1}), respectively:
\begin{equation}\label{RadialFunction2}
R_{nm}(\Phi;\rho)=\gamma_{\Phi;nm}\left[f_Y\left(\rho_1,\Lambda_{out}\right)J_{|m_\Phi|}\left(\pi\sqrt{E_{\Phi;nm}^\perp}\rho\right)-f_J\left(\rho_1,\Lambda_{out}\right)Y_{|m_\Phi |}\left(\pi\sqrt{E_{\Phi;nm}^\perp}\rho\right)\right] ,
\end{equation}
\begin{subequations}\label{FunctionsB}
\begin{eqnarray}\label{FunctionJ}
f_J\left(\rho,\Lambda\right)&=&\pi\sqrt{E_{\Phi;nm}^\perp}J'_{|m_\Phi |}\left(\pi\sqrt{E_{\Phi;nm}^\perp}\rho\right)+\frac{1}{\Lambda}J_{|m_\phi |}\left(\pi\sqrt{E_{\Phi;nm}^\perp}\rho\right),\\
\label{FunctionY}
f_Y\left(\rho,\Lambda\right)&=&\pi\sqrt{E_{\Phi;nm}^\perp}Y'_{|m_\Phi |}\left(\pi\sqrt{E_{\Phi;nm}^\perp}\rho\right)+\frac{1}{\Lambda}Y_{|m_\Phi |}\left(\pi\sqrt{E_{\Phi;nm}^\perp}\rho\right),
\end{eqnarray}
\end{subequations}
\begin{equation}\label{BesselEq1}
f_J\left(\rho_0,-\Lambda_{in}\right)f_Y\left(\rho_1,\Lambda_{out}\right)-f_Y\left(\rho_0,-\Lambda_{in}\right)f_J\left(\rho_1,\Lambda_{out}\right)=0,
\end{equation}
where $J_\nu(x)$ and $Y_\nu(x)$ are Bessel functions of the first kind of the order $\nu$ \cite{Abramowitz1}. As expected, in the limit of the large radius, $\rho_0\rightarrow\infty$, Eq.\ (\ref{BesselEq1}) transforms into the corresponding dependence of the straight waveguide \cite{Olendski2}:
\begin{equation}\label{StraightThresholds1}
\left(1+\frac{\Lambda_{in}}{\Lambda_{out}}\right)\cos\pi\sqrt{E_n^\perp}+\left(\frac{1}{\pi\sqrt{E_n^\perp}\Lambda_{out}}-\pi\sqrt{E_n^\perp}\Lambda_{in}\right)\sin\pi\sqrt{E_n^\perp}=0.
\end{equation}
Note that the azimuthal quantum number $m$ disappeared at the transition from (\ref{BesselEq1}) to (\ref{StraightThresholds1}) since for the straight duct only the transverse index $n$ defines the quantization.

So far, no use has been made of the complexity of the de Gennes distances: Eqs.\ (\ref{RadialFunction1}), (\ref{A1}) - (\ref{StraightThresholds1}) are valid for either real or complex parameters $\Lambda$. Miscellaneous cases of the real Robin parameter and different combinations of the background ${\bf B}$ and  the AB fields for the same 2D geometry have been addressed before \cite{Dalmasso1,Masale1,Ludu1,Zhu1,Zha1,Peshkin1,Halperin1,Skarzhinskiy1,Afanasev1,Makar1,Masale2,Groshev1,Avishai2,Makar2,Bruyndoncx1} in the framework of the formalism of linear equations (\ref{WaveEquation1}) or (\ref{Magnetic1}); in particular, Eq.\ (\ref{BesselEq1}) for the Dirichlet boundary conditions was presented in \cite{Peshkin1,Skarzhinskiy1,Afanasev1} with its uniform magnetic counterpart for zero AB flux given in \cite{Masale2,Avishai2} while Eq.\ (\ref{MagneticEq1}) without the AB field was derived in Refs. \cite{Masale1,Zhu1}.
 
From Eqs.\ (\ref{MagneticEq1}) and (\ref{BesselEq1}) it straightforwardly follows that for the complex extrapolation lengths the energies, in general,  are complex too,
\begin{equation}\label{ComplexEnergy1}
E_{\Phi;nm}^\perp=E_{\Phi;nm}^{(r)}-i\frac{\Gamma_{\Phi;nm}}{2}
\end{equation}
with real $E_{\Phi;nm}^{(r)}$ and $\Gamma_{\Phi;nm}$, and the following property holds:
\begin{equation}\label{conjugate1}
E_{\Phi;nm}^\perp\left(\overline{\Lambda}_{in},\overline{\Lambda}_{out}\right)=\overline{E_{\Phi;nm}^\perp\left(\Lambda_{in},\Lambda_{out}\right)}
\end{equation}
with the overline denoting a complex conjugate value. The same is true for the corresponding radial functions too. Physical meaning of the complex energy $E$ from Eq.\ (\ref{ComplexEnergy1}) was provided before \cite{Olendski1}; in particular, based on the standard quantum mechanical theory of scattering \cite{Newton1,Landau2}, it was stated that its imaginary part is inversely proportional to the lifetime $\tau$ of the corresponding quasi bound state:
\begin{equation}\label{lifetime1}
\tau_{\Phi;nm}=\frac{1}{\Gamma_{\Phi;nm}}.
\end{equation}
Also, since Eq.\ (\ref{RadialEquation1}) is invariant under the simultaneous transformations $m\rightarrow m\pm 1$, $\Phi\rightarrow\Phi\mp 1$, the following relations hold:
\begin{equation}
E_{\Phi;nm}=E_{\Phi\pm 1;n,m\mp 1},
\end{equation}
what is a general periodicity property of the AB systems \cite{Skarzhinskiy1,Afanasev1,Lewis1,Bagrov1}.

We repeat once again that even though (\ref{MagneticEq1}) and (\ref{BesselEq1}) were derived for the linearized GL equation, it can be shown that for the thin superconducting rings they follow also from the complete nonlinear GL theory. For doing this, one needs to employ the method of minimizing of the free energy $F$ of the superconducting state with respect to the coefficient linking the order parameter of the nonlinear equation with its linear counterpart, Eq.~(\ref{Factoring2}) \cite{Zhu1,Zha1}.

All other physical characteristics of the structure are found from the eigenfunctions $\Psi(\Phi;{\bf r})$ and eigenenergies $E_{\Phi;nm}$. For example, in the case of superconductors a magnetization operator
\begin{equation}\label{Magnetization1}
\hat{M}=i\frac{\partial}{\partial\varphi}-\frac{1}{2}B\rho^2
\end{equation}
is used to calculate the magnetic moment $M_z$ of the 2D ring:
\begin{equation}\label{Magnetization2}
M_z=\langle\psi_{nm}(\Phi;\rho,\varphi)|\hat{M}|\psi_{nm}(\Phi;\rho,\varphi)\rangle.
\end{equation}
One immediately gets:
\begin{equation}\label{Magnetization3}
M_z=-\left(m+\frac{B}{2}\int_{\rho_0}^{\rho_0+1}\left|R_{nm}(\Phi;\rho)\right|^2\rho^3 d\rho\right).
\end{equation}

An expression for the superconductor current density ${\bf j}_s$ in our dimensionless units is written as \cite{Ginzburg1,deGennes1}:
\begin{equation}\label{current1}
{\bf j}_s={\rm Im}\left[\overline{\Psi}\left(\bf r\right){\bm\nabla}\Psi\left(\bf r\right)\right]+{\bf A}\overline{\Psi}\left(\bf r\right)\Psi\left(\bf r\right).
\end{equation}
Similar formula (with, of course, ${\bf A}=0$) can be used for the Poynting vector in electrodynamics \cite{Jackson1} or for the sound energy density flux in acoustics \cite{Landau1}. For the 2D case without the $z$ dependence it transforms to
\begin{equation}\label{current2}
{\bf j}_s=\frac{1}{2\pi}\left[{\rm Im}\left(\overline{R}\frac{dR}{d\rho}\right){\bf e}_\rho+\left(\frac{m_\Phi}{\rho}+\frac{1}{2}B\rho\right)\left|R\right|^2{\bf e}_\varphi\right]
\end{equation}
with the unit orthogonal vectors ${\bf e}_\rho$ and ${\bf e}_\varphi$ along the radial and azimuthal directions, respectively. Since its divergence is proportional to $\Gamma$
\begin{equation}\label{divergence1}
{\rm div}{\bf j}_s=\frac{\pi}{4}|R|^2\Gamma,
\end{equation}
the positive (negative) imaginary part of the energy means that the corresponding spatial point serves as a sink (source) \cite{Olendski1}. In addition, Eq.\ (\ref{current2}) shows that the total current $J_\rho$ through the circle of the radius $\rho$ is:
\begin{equation}\label{current3}
J_\rho=\rho\int_0^{2\pi}d\varphi{j_s}_\rho=\rho{\rm Im}\left(\overline{R}\frac{dR}{d\rho}\right).
\end{equation}
In turn, for the 3D wire the total longitudinal supercurrent $J_z$ is given as
\begin{equation}\label{TotalZCurrent}
J_z=\int_{\rho_0}^{\rho_0+1}\rho d\rho\int_0^{2\pi}d\varphi {j_s}_z={\rm Re}(k_z)\exp\left[-2{\rm Im}(k_z)z\right].
\end{equation}
Eq.~(\ref{density1}) shows that the longitudinal dependence of the density $n_s$ is of the same form:
\begin{equation}\label{density2}
n_s=\left|R\left(\rho\right)\right|^2\exp\left[-2{\rm Im}(k_z)z\right].
\end{equation}
Expressions for ${\rm Re}(k_z)$ and ${\rm Im}(k_z)$ were derived and analyzed before \cite{Olendski1}, so, here we simply rewrite them pointing only to the fact that the imaginary part of the wave vector defining, according to (\ref{TotalZCurrent}), the longitudinal alteration of the current vanishes together with $\Gamma$:
\begin{subequations}\label{Wavevector3}
\begin{eqnarray}\label{Wavevector3Real}
{\rm Re}(k_z)&=&\frac{\pi}{\sqrt{2}}\sqrt{\sqrt{\left[E-E_{\Phi;nm}^{\left(r\right)}\right]^2+\left(\Gamma_{\Phi;nm}/2\right)^2}+\left[E-E_{\Phi;nm}^{\left(r\right)}\right]}\\
\label{Wavevector3Imag}
{\rm Im}(k_z)&=&\frac{\pi}{\sqrt{2}}\frac{\Gamma_{\Phi;nm}/2}{\sqrt{\sqrt{\left[E-E_{\Phi;nm}^{\left(r\right)}\right]^2+\left(\Gamma_{\Phi;nm}/2\right)^2}+\left[E-E_{\Phi;nm}^{\left(r\right)}\right]}}.
\end{eqnarray}
\end{subequations}
In other words, the lossless longitudinal flow is achieved when the total transverse energy is real, as expected. As it will be shown below, for the multiply connected geometry it takes place when the total 2D radial current through the inner circle $J_{\rho_0}$ is equal to its counterpart through the outer border $J_{\rho_1}$.

\section{Results and discussion}
\label{sec3}
In this section the outcome of the calculations based on the theory developed in Chapter \ref{sec2} is presented and their mathematical and physical interpretations are given. If in the results below the index or subscript $\Phi$ is dropped, it means that its particular value of $\Phi=0$ is used; otherwise, it takes any arbitrary magnitudes.

\subsection{Columnar defect}\label{ColumnarDefect1}
Before discussing the annular geometry with the {\em finite} inner and outer radii, it is instructive to consider the case of the cavity in the otherwise uniform material. In terms of Fig.~\ref{Fig1} this corresponds to the infinite width $d$ of the ring, $d\rightarrow\infty$. The only remaining distance $\rho_0$ obviously becomes the natural unit of length; accordingly, all other physical quantities containing distances will be correspondingly rescaled. This convention will be assumed throughout the whole present subsection for either real or complex Robin parameters. As there is only one surface, in the following the subscript 'in' at the extrapolation length $\Lambda$ will be dropped. For our treatment, such a geometry presents an interest since it is an asymptotic case of the strong magnetic field that pushes the particle closer to the origin and, thus, the influence of the outer cylindrical surface becomes negligible. However, properties of the system with the hole in the film are important by themself with possible application, first of all, in superconductivity and they were addressed by the number of theoretical \cite{Bruyndoncx1,Buzdin1,Ovchinnikov1,Bezryadin1,Bezryadin2,Bezryadin3,Bruyndoncx2,Melnikov1} and experimental \cite{Bezryadin1,Bezryadin3,Bezryadin4,Gutierrez1} studies limited, however, to the real de Gennes distances with its, primarily, infinite value (Neumann boundary condition). Here, we will present the results that escaped an attention of the previous researchers. We will also make a comparative analysis with its inverse counterpart of the solid cylinder. Below, the geometry  of the columnar defect will be called the exterior configuration, and the confined disk - the interior one \cite{Hornberger1,Hornberger2}.

Obviously, for the antidot with unrestricted at infinity motion of the particle, the  bound states for the 2D geometry exist only when the background magnetic field is {\em not} zero, ${\bf B}\ne {\bf 0}$. Assuming the same gauge for the vector potential as before and taking into account asymptotic properties of the confluent hypergeometric functions at infinity \cite{Abramowitz1,Bateman1}, one writes the following expression for the radial part of the total wave function:
\begin{equation}\label{PillarFunction1}
R_{nm}(\Phi;\rho)=\gamma_{\Phi;nm}\exp\left(-\frac{B}{4}\rho^2\right)\left(\frac{B}{2}\rho^2\right)^{|m_\Phi|/2}U\left(\frac{m_\Phi+|m_\Phi|+1}{2}-\frac{\pi^2}{2}\frac{E_{\Phi;nm}^\perp}{B},|m_\Phi|+1,\frac{B}{2}\rho^2\right).
\end{equation}
Robin demand at the boundary $\rho=1$ leads to the following transcendental equation for the determination of the energies $E_{\Phi;nm}^\perp$:
\begin{eqnarray}
&&\left(\left|m_\Phi\right|-\frac{B}{2}-\frac{1}{\Lambda}\right)U\left(\frac{m_\Phi+|m_\Phi|+1}{2}-\frac{\pi^2}{2}\frac{E_{\Phi;nm}^\perp}{B},\left|m_\Phi\right|+1,\frac{B}{2}\right)\nonumber\\
\label{PillarEq1}
&&+BU'\left(\frac{m_\Phi+|m_\Phi|+1}{2}-\frac{\pi^2}{2}\frac{E_{\Phi;nm}^\perp}{B},\left|m_\Phi\right|+1,\frac{B}{2}\right)=0,
\end{eqnarray}
where, instead of the derivative of the function $U$, one can use the confluent hypergeometric function itself with the different parameters according to \cite{Abramowitz1}:
$$
U'(a,b,z)=-aU(a+1,b+1,z).
$$
Eq.\ (\ref{PillarEq1}) for $\Phi=0$ was derived before \cite{Bezryadin3} (see also Ref. \cite{Bruyndoncx2} for the Neumann limit). Normalization condition, Eq.~(\ref{normalization1}), turns to
\begin{equation}\label{normalization3}
\int_1^\infty\left|R_{nm}(\Phi;\rho)\right|^2\rho d\rho=1,
\end{equation}
and expression for the magnetization reads:
\begin{equation}\label{Magnetization4}
M_z=-\left(m+\frac{B}{2}\int_1^\infty\left|R_{nm}(\Phi;\rho)\right|^2\rho^3 d\rho\right).
\end{equation}
Four numbered equations above are valid for the real as well as complex $\Lambda$.

Using properties of the function $U(a,b,x)$ \cite{Abramowitz1}, one immediately sees from Eq.\ (\ref{PillarEq1}) that for the {\em small} magnetic field, $B\rightarrow 0$, the energy spectrum transforms into the Landau levels  \cite{Page1,Landau2} disturbed by the AB flux \cite{Skarzhinskiy1,Afanasev1,Lewis1,Bagrov1}:
\begin{equation}\label{LandauLevels1}
E_{\Phi;nm}^\perp=\frac{2}{\pi^2}\left(n+\frac{m_\Phi+\left|m_\Phi\right|+1}{2}\right)B.
\end{equation}
This is in a sharp contrast with the interior problem where the limit of Eq. (\ref{LandauLevels1}) is achieved for the {\em large} intensities $B$ \cite{Olendski1}. In the latter case, the strong fields push the charged carrier closer to the center of the disk and, due to the small magnetic radius $r_B=B^{-1/2}$, the walls of the dot only slightly perturb the Landau-AB states, Eq. (\ref{LandauLevels1}). In turn, for the columnar defect in the small fields, the cavity radius is much smaller than the diverging distance $r_B$ and, thus, the particle with energy from Eq. (\ref{LandauLevels1}) almost does not 'see' the antidot and does not 'feel' the perturbation caused by it.  

\subsubsection{Real de Gennes distance}
Another asymptotic limit of Eq. (\ref{PillarEq1}) can be analytically derived for the small nonzero magnetic fields and the extrapolation length tending to zero from the left, $\Lambda\rightarrow -0$. Then, one gets for the principal eigenvalues, $n=0$, of each fixed $m$:
\begin{equation}\label{PillarEq2}
E_{\Phi;0,m}^\perp=-\frac{1}{\pi^2\Lambda^2}, \ \Lambda\rightarrow -0, \ B\rightarrow 0.
\end{equation}
The same dependence is obtained from equation
\begin{equation}\label{RadialRobin1}
\pi\sqrt{|E_{\Phi;0,m}^\perp|}I_{|m_\Phi|}'\left(\pi\sqrt{|E_{\Phi;0,m}^\perp|}\right)+\frac{1}{\Lambda}I_{\left|m_\Phi\right|}\left(\pi\sqrt{|E_{\Phi;0,m}^\perp|}\right)=0,
\end{equation}
describing the motion of the particle with negative energy $E_{\Phi;0,m}^\perp$ inside the field-free disk with negative extrapolation length $\Lambda$ \cite{Olendski2}. Here, $I_\nu(x)$ is a modified Bessel function of the order $\nu$ \cite{Abramowitz1}. In the same way, for the annulus it can be shown from Eq. (\ref{BesselEq1}) that for the two (generally different) infinitely small negative extrapolation lengths $\Lambda_{in,out}$ the two lowest energies $E_{\Phi;0,m}$ and $E_{\Phi;1,m}$ for each $m$ are:
\begin{equation}\label{RadialRobin2}
E_{\Phi;\left(0,1\right),m}^\perp=-\frac{1}{\pi^2\Lambda_{in,out}^2},\quad \Lambda_{in}\rightarrow -0, \Lambda_{out}\rightarrow -0.
\end{equation}
In fact, results (\ref{PillarEq2}) and (\ref{RadialRobin2}) are a particular case of more general property of the vanishingly small negative Robin parameter intensively studied recently by mathematicians \cite{Lacey1,Lou1,Levitin1,Daners1,Colorado1} (see also \cite{Balian1,Berry2}). Systems with negative extrapolation lengths raise interesting theoretical questions about fundamentals of quantum mechanics; for example, for an arbitrarily shaped domain with general perfectly reflecting walls a generalized Heisenberg uncertainty relation was derived earlier this year \cite{AlHashimi1}. As it was mentioned in the Introduction, experimentally such systems have been fabricated with the help of superconductors \cite{Fink1,Kozhevnikov1,Kozhevnikov2}. According to Eqs.\ (\ref{TotalExtrapolationLength1}) and (\ref{PotentialU}), the limit of the negligibly small negative de Gennes distance can be achieved by applying to such structures appropriately directed electric field $\cal E$ when the total extrapolation length $\Lambda_{tot}$ approaches zero from the left.
\begin{figure}
\centering
\includegraphics[width=0.95\columnwidth]{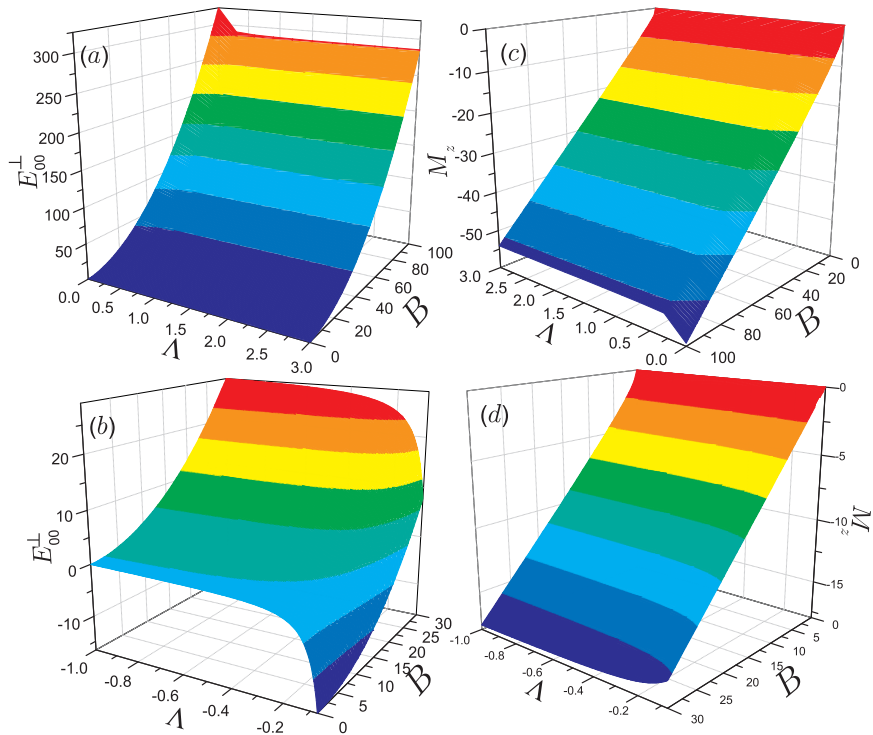}
\caption{\label{Fig2}
Energy $E_{00}^\perp$ (left panels) and corresponding magnetization $M_z$ (right panels) of the columnar defect as functions of magnetic field $B$ and positive (upper panels) and negative (lower panels) extrapolation length $\Lambda$. Note different scales for the upper and lower panels. To better emphasize the obtained dependencies, the $\Lambda$- and $B$- axes in panel (c) have been reversed as compared to panel (a). For the same reason, the field axis in panel (d) grows in the opposite direction of panel (b).
}
\end{figure}

Fig.~\ref{Fig2} shows $E_{00}^\perp$ as a function of the positive [panel (a)] and negative [panel (b)] de Gennes distance $\Lambda$ and magnetic intensity $B$. It is seen that the energy monotonically increases with the field for all extrapolation lengths. The asymptotic limit of the function $U(a,b,x)$ for the large positive $x$ and negative $a$ such that $x<2b-4a$ \cite{Abramowitz1,Buchholz1}\footnote{Equations (13.5.21) and (13.5.22) in Ref. \cite{Abramowitz1} are wrong. The correct expressions can be obtained from Eqs. (8.11) and (8.10) of Ref. \cite{Buchholz1} for the Whittaker functions $M_{\kappa,\mu}(x)$ and $W_{\kappa,\mu}(x)$ and their relations to the functions $M(a,b,x)$ and $U(a,b,x)$.}
\begin{equation}
U(a,b,x)=2\exp\left[\kappa\ln\left(\frac{\kappa}{e}\right)+\frac{x}{2}\right]x^{-b/2}\frac{1}{\sqrt{\tan\theta}}\sin\left[\kappa\left(2\theta-\sin 2\theta\right)+\frac{\pi}{4}\right]
\end{equation}
with $\kappa=b/2-a$ and $\cos^2\theta=x/(4\kappa)$, leads, for the Dirichlet case, $\Lambda=0$, to the following expression of the energies $E_{\Phi;nm}^\perp$ in the strong fields:
\begin{equation}\label{PillarEq3}
\left.E_{\Phi;nm}^\perp\right|_{\Lambda=0}=\frac{1}{4\pi^2}\left\{1+\left[\left(n+\frac{3}{4}\right)\frac{6\pi}{B}\right]^{2/3}\right\}B^2+\frac{m_\Phi}{\pi^2}B, \quad B\rightarrow\infty.
\end{equation}
As Fig.~\ref{Fig2} demonstrates, the interplay between the magnetic field and the Robin electrostatic potential leads to the about the same quadratic dependence at the strong fields for all extrapolation lengths while the $\Lambda$-dependence in the same limit is noticeable for the small positive Robin distances only when $1/\Lambda$ in Eq. (\ref{MagneticEq1}) is comparable to $B/2$, and for the larger magnitudes of the de Gennes distance the energy $E_{\Phi;nm}^\perp$ is almost $\Lambda$-independent.

The onset of the peculiarity from Eq.~(\ref{PillarEq2}) is clearly seen in panel (b). It also shows that with increasing intensity $B$ the negative energy grows too, turns to zero at some field and unrestrictedly increases in the positive direction with further growth of $B$. To clarify the interaction between the magnetic field and negative Robin parameter, it is instructive to consider the extrapolation length $\Lambda_{\Phi;m}^{\left(0\right)}$ at which the energy is zero, $E_{\Phi;0,m}^\perp=0$. It directly follows from Eq.~(\ref{PillarEq1}) that it is
\begin{equation}\label{LambdaZeroEnergy1}
\frac{1}{\Lambda_{\Phi;m}^{\left(0\right)}}=\left|m_\Phi\right|-\frac{B}{2}\left[1+\left(m_\Phi+\left|m_\Phi\right|+1\right)\frac{U\left(\frac{m_\Phi+\left|m_\Phi\right|+1}{2}+1,\left|m_\Phi\right|+2,\frac{B}{2}\right)}{U\left(\frac{m_\Phi+\left|m_\Phi\right|+1}{2},\left|m_\Phi\right|+1,\frac{B}{2}\right)}\right].
\end{equation}
Utilizing asymptotic properties of the confluent hypergeometric function $U(a,b,x)$\footnote{We point out other typos in the reference literature; namely, Eqs. (13.5.9) in \cite{Abramowitz1} and (6.8.5) in \cite{Bateman1} are wrong. For the correct form, one needs in Eq. (6.8.5) of Ref. \cite{Bateman1} to convert the rightmost negative sign into its positive counterpart.}, one gets in the limiting cases of the small and strong fields:
\begin{equation}\label{LambdaZeroEnergy2}
\frac{1}{\Lambda_{\Phi;m}^{\left(0\right)}}=\left\{
\begin{array}{cc}
\left.\begin{array}{cc}
\frac{2}{\ln\left(B/8\right)+\gamma}-\frac{B}{2},&m_\Phi =0\\
-\left(\left|m_\Phi\right|+\left. B\right/2\right),&m_\Phi\ne 0
\end{array}
\right\}&B\ll 1\\
-\left(B/2+m_\Phi+1\right),&B\gg 1.
\end{array}
\right.
\end{equation}
Here, $\gamma=\lim_{n\rightarrow\infty}\left(\sum_{k=1}^n\frac{1}{k}-\ln n\right)=0.57721\ldots$ is Euler's constant \cite{Abramowitz1}.
\begin{figure}
\centering
\includegraphics[width=0.99\columnwidth]{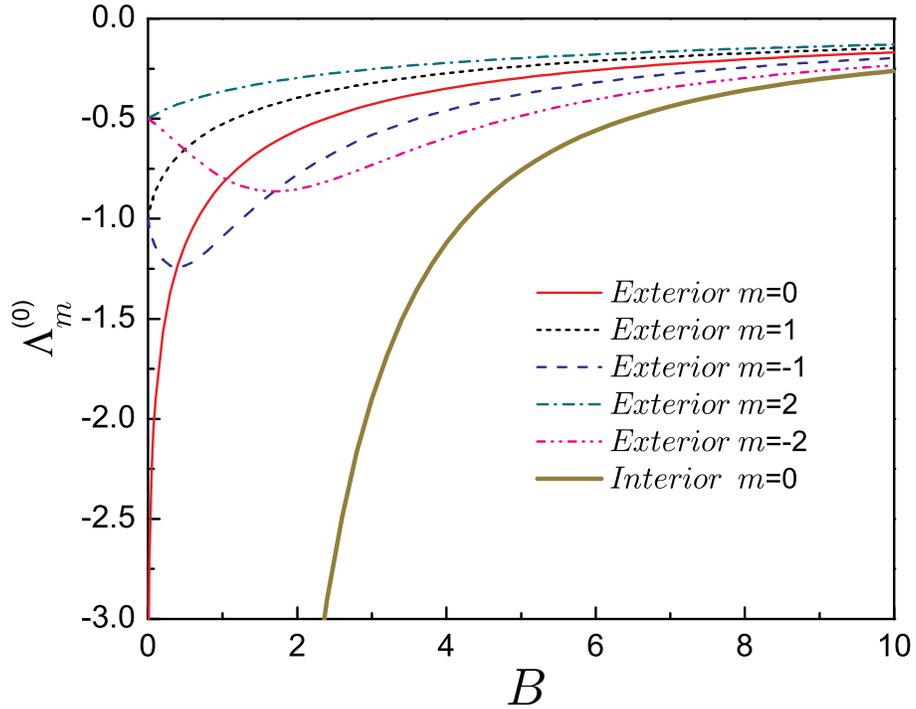}
\caption{\label{Fig3}
Critical extrapolation length $\Lambda_m^{\left(0\right)}$ of the columnar defect as a function of the magnetic field $B$ where the solid line is for $m=0$, dotted line - for $m=1$, dashed line - for $m=-1$, dash-dotted curve - for $m=2$, and dash-dot-dotted curve - for $m=-2$. For comparison, $\Lambda_0^{\left(0\right)}$ of the quantum disk is also shown by the thick solid line.
}
\end{figure}

Fig.~\ref{Fig3} shows $\Lambda_{m}^{\left(0\right)}$ dependence on $B$ for several quantum numbers $m$. The states with $\Lambda$ lying above (below) the corresponding curves possess negative (positive) energies. Increasing field shrinks the range of extrapolation lengths with the energies lying below zero: for the smaller magnitudes of the de Gennes distance the larger magnetic intensities are needed to pull out the energy into the positive area. For comparison, we also plot in Fig.~\ref{Fig3} the critical extrapolation length of the dot calculated from \cite{Olendski1}
\begin{equation}\label{LambdaZeroEnergy3}
\frac{1}{\Lambda_{\Phi;m}^{\left(0\right)}}=-\left|m_\Phi\right|+\frac{B}{2}\left[1-\frac{m_\Phi+\left|m_\Phi\right|+1}{|m_\Phi|+1}\frac{M\left(\frac{m_\Phi+\left|m_\Phi\right|+1}{2}+1,\left|m_\Phi\right|+2,\frac{B}{2}\right)}{M\left(\frac{m_\Phi+\left|m_\Phi\right|+1}{2},\left|m_\Phi\right|+1,\frac{B}{2}\right)}\right].
\end{equation}
Eq.~(\ref{LambdaZeroEnergy2}) for the interior problem reads:
\begin{equation}\label{LambdaZeroEnergy4}
\frac{1}{\Lambda_{\Phi;m}^{\left(0\right)}}=\left\{
\begin{array}{cc}
\left.\begin{array}{cc}
-B^2/16,&m_\Phi =0\\
-\left(\left|m_\Phi\right|+\frac{m_\Phi}{\left|m_\Phi\right|+1}\left. B\right/2\right),&m_\Phi\ne 0
\end{array}
\right\}&B\ll 1\\
-\left(B/2+\left|m_\Phi\right|\right),&B\gg 1.
\end{array}
\right.
\end{equation}
Comparison of Eqs.~(\ref{LambdaZeroEnergy2}) and (\ref{LambdaZeroEnergy4}) shows that for the principal state, $m_\Phi=0$, much larger magnetic fields are needed for the disk geometry to reach a positive energy. This can be understood after consideration of the factors which affect energy behavior. For either case, tending to zero negative Robin length pushes energy downwards. For the quantum dot, an increasing magnetic field is the only parameter that forces the energy to grow. For the columnar defect it is additionally aided by the cylindrical electrostatic potential of the hole. Combined effective potential of the magnetic squeezing and electric confinement gets narrower for the larger $B$, and so, their mutual effort in withstanding against the opposite trend of the de Gennes distance requires smaller fields to see the energy $E_{\Phi;0,m}$ positive.

Right panels of Fig.\ \ref{Fig2} show magnetizations $M_z$ corresponding to the energy of the state depicted in the left parts. 
Its dependence on the field shows completely different behavior as compared to the interior problem (see, e.g., Fig. 9 in Ref. \cite{Olendski1}). It is known \cite{White1} that the magnetic moment and the {\em real} energy $E$ are related as
\begin{equation}\label{MagnetEnergy1}
{\bf M}_z=-\pi^2\frac{\partial E}{\partial {\bf B}}.
\end{equation}
Accordingly, at the strong intensities, the magnitude of $M_z$ for the Dirichlet case, $\Lambda=0$, grows almost linearly with the field:
\begin{equation}\label{PillarEq4}
\left.M_z\right|_{\Lambda=0}=
-\left(m_\Phi+\left\{1+\left[\left(n+\frac{3}{4}\right)\frac{6\pi}{B}\right]^{2/3}\right\}\frac{B}{2}
-\left[6\pi\left(n+\frac{3}{4}\right)\right]^{2/3}\frac{B^{1/3}}{6}\right),\ B\rightarrow\infty,
\end{equation}
as it follows from Eq.\ (\ref{PillarEq3}). Similar to the energy, the $\Lambda$ dependence of magnetization is noticeable in this regime for the small Robin parameters only. For the small negative de Gennes distance, $\Lambda\rightarrow -0$, the lowest state, $n=0$, of each azimuthal quantum number $m$ is localized near the interface with the radial component of its wave function closer and closer resembling the $\delta$-function; accordingly, the magnetization turns to
\begin{equation}\label{PillarEq5}
M_z=-\left(m+\frac{B}{2}\right),\quad \Lambda\rightarrow -0.
\end{equation}

\subsubsection{Complex Robin parameter}\label{Section3_1_2}
It was shown in Ref.~\cite{Olendski1} that nonzero real part of the complex extrapolation length $\Lambda$ quenches the resonant features of the system parameters dependence on the imaginary component of $\Lambda$. As the same is true for the Robin antidot too, below we will present the results for the purely imaginary Robin parameter $\Lambda\equiv i\lambda$.

\begin{figure}
\centering
\includegraphics[width=0.99\columnwidth]{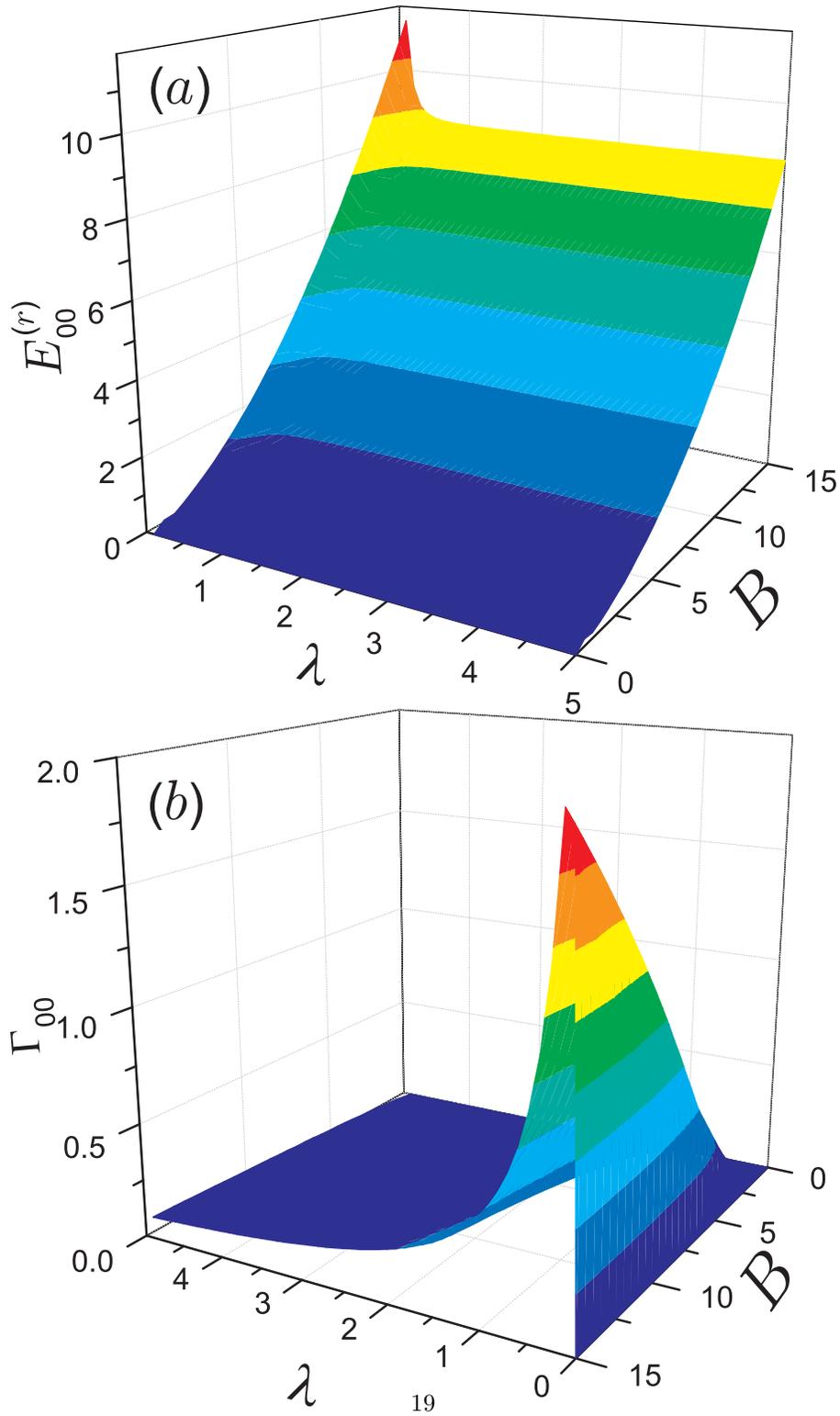}
\caption{\label{Fig4}
(a) Real $E_{00}^{(r)}$ and (b) negative double imaginary $\Gamma_{00}$ parts of the total transverse energy $E_{00}^\perp$ as functions of $\lambda$ and $B$ for the columnar defect. Note reversed $\lambda$- and $B$-axes directions in panels (a) and (b).
}
\end{figure}
Fig.\ \ref{Fig4} depicts $E_{00}^{(r)}$ and $\Gamma_{00}$ as functions of $\lambda$ and $B$. The dependence of the real parts of the total energy on the field and on the complex Robin parameter is quite similar to that of the system with positive real extrapolation length [cf. panels (a) of Figs.\ \ref{Fig2} and \ref{Fig4}]; namely, in either case, they almost quadratically depend on the magnetic intensity in the high-field regime while the magnitude of the Robin parameter $|\Lambda|$ alters the energy at high fields only when it is of the order $|\Lambda|\sim 2/B$. In turn, the imaginary part exhibits a sharp extremum as a function of the purely imaginary de Gennes distance. The same resonant dependence was observed for the interior problem too \cite{Olendski1}. The origins of these resonances are of the same nature: systems with almost purely Dirichlet case at $|\lambda|\ll 1$ and that close to its Neumann counterpart, $|\lambda|\gg 1$, are characterized by the same sign of $\Gamma$ and its linear dependence on $|\lambda|$ and $1/|\lambda|$, respectively; accordingly, interaction of these two asymptotics in the intermediate regime, $|\lambda|\sim 1$, leads to a pronounced maximum with its magnitude being $n$ and $m$ dependent. The drastic difference between the exterior and interior configurations lies in the $\Gamma$-$B$ dependence: if, for the quantum disk, the field quenches the resonance, then for the hole, as panel (b) of Fig. \ref{Fig4} vividly demonstrates, the extremum increases with $B$. This inverse behavior is explained by the localization properties of the magnetic field: its growth, in the case of the quantum dot, pushes the charged carrier closer to the axis and, thus, the influence of the Robin potential on the properties of the system  diminishes \cite{Olendski1}, while for the columnar defect it moves the particle closer to the circular impenetrable barrier and in this way increases scattering. Calculations show that the maximum of the resonance for the strong $B$ almost quadratically depends on the field what is consistent with the result of the zero de Gennes distance, Eq.\ (\ref{PillarEq3}). Another difference between the two configurations is explained by the similar reasoning; namely, for the interior disk the resonance increases with the radial $n$ and azimuthal $|m|$ quantum numbers \cite{Olendski1} and for the perforated film at the fixed field the opposite is true since the radial wave function for the larger $n$ and $|m|$ is spread out wider from the system origin and, thus, its deformation by the hole (disk) is smaller (larger).

\subsection{Annulus with complex extrapolation lengths}
Similar to subsection \ref{Section3_1_2}, our analysis of the hollow waveguide will concentrate on the purely imaginary Robin parameters: $\Lambda_{in,out}\equiv i\lambda_{in,out}$, since in this case the obtained energy and current dependencies on $\Lambda$ are the most conspicuous ones. First, the results are discussed for $B=0$ with the emphasis on the asymptotic cases and the conditions under which the energy $E_{\Phi;nm}^\perp$ is real. Influence of the background uniform magnetic field $\bf B$ will be presented separately in the second subsection.

\subsubsection{Field-free case}
We start our analysis from the study of the asymptotic cases of the large and small $|\lambda|$; namely, similar to the solid cylinder case \cite{Olendski1}, the limiting cases of the small, $|\lambda_{in,out}|\ll 1$, and large, $|\lambda_{in,out}|\gg 1$, magnitudes of the purely imaginary de Gennes distances can be derived from Eq.\ (\ref{BesselEq1}) using the addition theorem for the Bessel functions and their asymptotic properties \cite{Abramowitz1}. Expressions for the imaginary part of the energy $\Gamma_{\Phi;nm}$ are provided below for all possible permutations of the asymptotics:
\begin{subequations}\label{Asymptotics1}
\begin{eqnarray}\label{Asymptotics1DD}
\Gamma_{\Phi;nm}=-\frac{4}{\pi^2}\left(x_{m_\Phi n}^{DD}\right)^2\frac{\lambda_{in}{\cal F}_{m_\Phi}^{ND}\left(x_{m_\Phi n}^{DD}\right)-\lambda_{out}{\cal F}_{m_\Phi}^{DN}\left(x_{m_\Phi n}^{DD}\right)}{\rho_0{\cal F}_{m_\Phi}^{ND}\left(x_{m_\Phi n}^{DD}\right)-\rho_1{\cal F}_{m_\Phi}^{DN}\left(x_{m_\Phi n}^{DD}\right)},\quad\left|\lambda_{in,out}\right|\ll 1\qquad\qquad\\
\label{Asymptotics1DN}
\Gamma_{\Phi;nm}=-\frac{4}{\pi^2}\frac{\left(x_{m_\Phi n}^{DN}\right)^2\lambda_{in}{\cal F}_{m_\Phi}^{NN}\left(x_{m_\Phi n}^{DN}\right)+\left.{\cal F}_{m_\Phi}^{DD}\left(x_{m_\Phi n}^{DN}\right)\right/\lambda_{out}}{\rho_0{\cal F}_{m_\Phi}^{NN}\left(x_{m_\Phi n}^{DN}\right)+\left[\frac{\left|m_\Phi\right|^2}{\left(x_{m_\Phi n}^{DN}\rho_1\right)^2}-1\right]\rho_1{\cal F}_{m_\Phi}^{DD}\left(x_{m_\Phi n}^{DN}\right)},\quad\left|\lambda_{in}\right|\ll 1,\left|\lambda_{out}\right|\gg 1\qquad\\
\label{Asymptotics1ND}
\Gamma_{\Phi;nm}=-\frac{4}{\pi^2}\frac{\left.{\cal F}_{m_\Phi}^{DD}\left(x_{m_\Phi n}^{ND}\right)\right/\lambda_{in}-\left(x_{m_\Phi n}^{ND}\right)^2\lambda_{out}{\cal F}_{m_\Phi}^{NN}\left(x_{m_\Phi n}^{ND}\right)}{-\left[\frac{\left|m_\Phi\right|^2}{\left(x_{m_\Phi n}^{DN}\rho_0\right)^2}-1\right]\rho_0{\cal F}_{m_\Phi}^{DD}\left(x_{m_\Phi n}^{ND}\right)+\rho_1{\cal F}_{m_\Phi}^{NN}\left(x_{m_\Phi n}^{ND}\right)},\quad\left|\lambda_{in}\right|\gg 1,\left|\lambda_{out}\right|\ll 1\qquad\\
\Gamma_{\Phi;nm}=-\frac{4}{\pi^2}\left\{\frac{-\left.{\cal F}_{m_\Phi}^{DN}\left(x_{m_\Phi n}^{NN}\right)\right/\lambda_{in}+\left.{\cal F}_{m_\Phi}^{ND}\left(x_{m_\Phi n}^{NN}\right)\right/\lambda_{out}}{\left[\frac{\left|m_\Phi\right|^2}{\left(x_{m_\Phi n}^{NN}\rho_0\right)^2}-1\right]\rho_0{\cal F}_{m_\Phi}^{DN}\left(x_{m_\Phi n}^{NN}\right)+\left[\frac{\left|m_\Phi\right|^2}{\left(x_{m_\Phi n}^{NN}\rho_1\right)^2}-1\right]\rho_1{\cal F}_{m_\Phi}^{ND}\left(x_{m_\Phi n}^{NN}\right)}\left(1-\delta_{\left|m_\Phi\right|+n,0}\right)\right.\nonumber\\
\label{Asymptotics1NN}
\left.-\frac{1}{2\rho_0+1}\left(\frac{\rho_0}{\lambda_{in}}+\frac{\rho_1}{\lambda_{out}}\right)\delta_{m_\Phi,0}\delta_{n,0}
\right\},\quad\left|\lambda_{in,out}\right|\gg 1.\quad
\end{eqnarray}
\end{subequations}
Here, $\delta_{\nu\nu'}=\left\{\begin{array}{cc}1&\nu=\nu'\\0&\nu\ne\nu'\end{array}\right.$ is a Kronnecker symbol for, in general, real $\nu$ and $\nu'$, and
\begin{subequations}\label{Asymptotics2}
\begin{eqnarray}\label{Asymptotics2DD}
{\cal F}_\nu^{DD}(x)&=&J_{\left|\nu\right|}(\rho_0x)Y_{\left|\nu\right|}(\rho_1x)-Y_{\left|\nu\right|}(\rho_0x)J_{\left|\nu\right|}(\rho_1x)\\
\label{Asymptotics2DN}
{\cal F}_\nu^{DN}(x)&=&J_{\left|\nu\right|}(\rho_0x)Y_{\left|\nu\right|}'(\rho_1x)-Y_{\left|\nu\right|}(\rho_0x)J_{\left|\nu\right|}'(\rho_1x)\\
\label{Asymptotics2ND}
{\cal F}_\nu^{ND}(x)&=&J_{\left|\nu\right|}'(\rho_0x)Y_{\left|\nu\right|}(\rho_1x)-Y_{\left|\nu\right|}'(\rho_0x)J_{\left|\nu\right|}(\rho_1x)\\
\label{Asymptotics2NN}
{\cal F}_\nu^{NN}(x)&=&J_{\left|\nu\right|}'(\rho_0x)Y_{\left|\nu\right|}'(\rho_1x)-Y_{\left|\nu\right|}'(\rho_0x)J_{\left|\nu\right|}'(\rho_1x)
\end{eqnarray}
\end{subequations}
and $x_{\nu n}^{ij}$ with $i$, $j$ running over the indices $D$ and $N$ is $n$th solution of the equation zeroing the corresponding ${\cal F}_\nu^{ij}(x)$:
\begin{equation}\label{x_mn}
{\cal F}_\nu^{ij}(x_{\nu n}^{ij})=0,\quad i,j=D\ {\rm or}\ N
\end{equation}
with the assumption that $x_{00}^{NN}=0$. In other words, $x_{\nu n}^{ij}$ are the eigenvalues of the 2D boundary problem with the Dirichlet ($D$) and/or Neumann ($N$) requirements at the corresponding edge of the annulus with the first (second) character of the superscript denoting inner (outer) circle. Their dependence on the radius $\rho_0$ for $\nu=0$ was analysed before \cite{Olendski4}. In the asymptotic limit of the large radius, $\rho_0\rightarrow\infty$, Eqs.\ (\ref{Asymptotics1}) transform to
\begin{subequations}\label{AsymtoticsStraight1}
\begin{eqnarray}
\label{AsymtoticsStraight1DD}
\Gamma_n&=&4\left(n+1\right)^2\left(\lambda_{in}+\lambda_{out}\right), \quad\left|\lambda_{in,out}\right|\ll 1\\
\label{AsymtoticsStraight1Mixed}
\Gamma_n&=&\frac{4}{\pi^2}\left[1/\lambda_>+\pi^2\left(n+1/2\right)^2\lambda_<\right],\quad\left|\lambda_<\right|\ll 1,\left|\lambda_>\right|\gg 1\\
\label{AsymtoticsStraight1NN}
\Gamma_n&=&\frac{4}{\pi^2}\frac{1}{1+\delta_{n0}}\left(\frac{1}{\lambda_{in}}+\frac{1}{\lambda_{out}}\right),\quad\left|\lambda_{in,out}\right|\gg 1
\end{eqnarray}
\end{subequations}
derivable, of course, also from Eq.\ (\ref{StraightThresholds1}). In Eq.\ (\ref{AsymtoticsStraight1Mixed}) the lengths $\lambda_<$ and $\lambda_>$ are defined according to $\left|\lambda_<\right|=\min\left(\left|\lambda_{in}\right|,\left|\lambda_{out}\right|\right)$, $\left|\lambda_>\right|=\max\left(\left|\lambda_{in}\right|,\left|\lambda_{out}\right|\right)$.
\begin{figure}
\centering
\includegraphics[width=0.99\columnwidth]{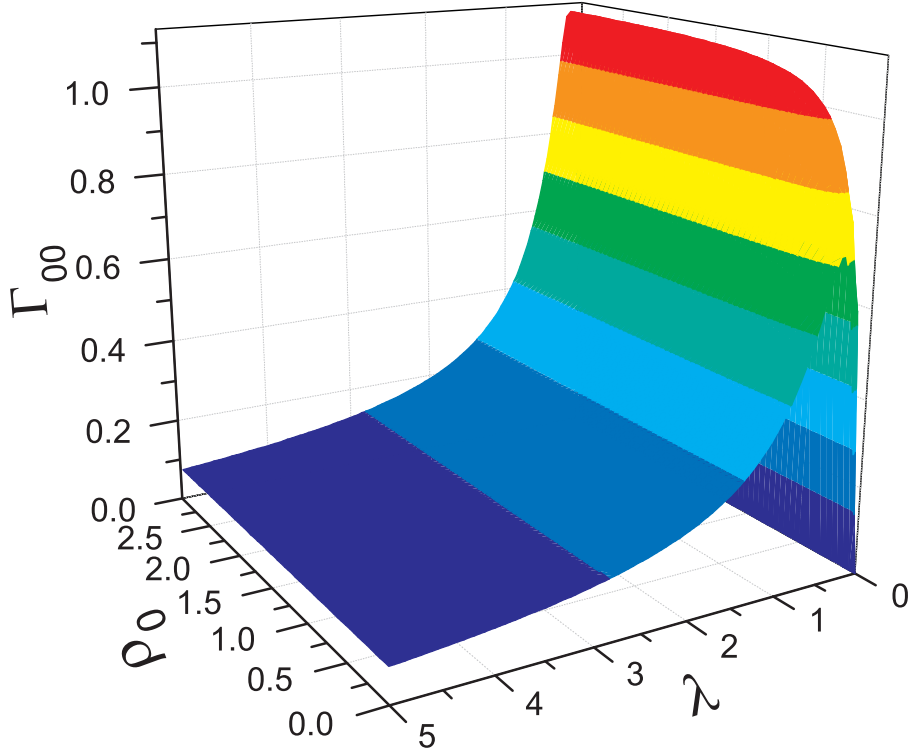}
\caption{\label{Fig5}
Negative double imaginary part $\Gamma_{00}$ of the total transverse energy $E_{00}^\perp$ as a function of the radius $\rho_0$ and equal inner and outer purely imaginary distances $\lambda=\lambda_{in}=\lambda_{out}$.
}
\end{figure}

Fig.~\ref{Fig5} demonstrates a transformation of the imaginary part of the energy with changing inner radius and equal outer and inner extrapolation lengths, $\lambda=\lambda_{in}=\lambda_{out}$. It is seen that with increasing the radius $\rho_0$ the magnitude of the resonance grows and, for $\rho_0\gtrsim 2$ it saturates to its value of the straight waveguide. Larger scattering for the straight channel can be seen from the comparison between Eq.\ (\ref{AsymtoticsStraight1DD}) and the corresponding expression $\Gamma_{\Phi;nm}=4j_{|m_\Phi|n}^2\lambda/\pi^2$ for the solid cylinder at the small $|\lambda|$ \cite{Olendski1}. Here, $j_{\nu n}$ is $n$th root of equation $J_\nu(x)=0$ \cite{Abramowitz1}. In this regime, for $m=n=\Phi=0$ the imaginary part of the energy for the straight channel is about 3.41 times larger. Physically, this difference is explained by the centrifugal forces acting in the curved sample \cite{Nesvizhevsky1} and being absent for the straight film where both interfaces contribute equally to the reactive scattering while for the annulus the unequal distribution of the wave function along the radius leads to the increased influence of one boundary and the decreased contribution from the other surface with their total mutual effort being smaller than for the unbent structure.
\begin{figure}
\centering
\includegraphics[width=0.99\columnwidth]{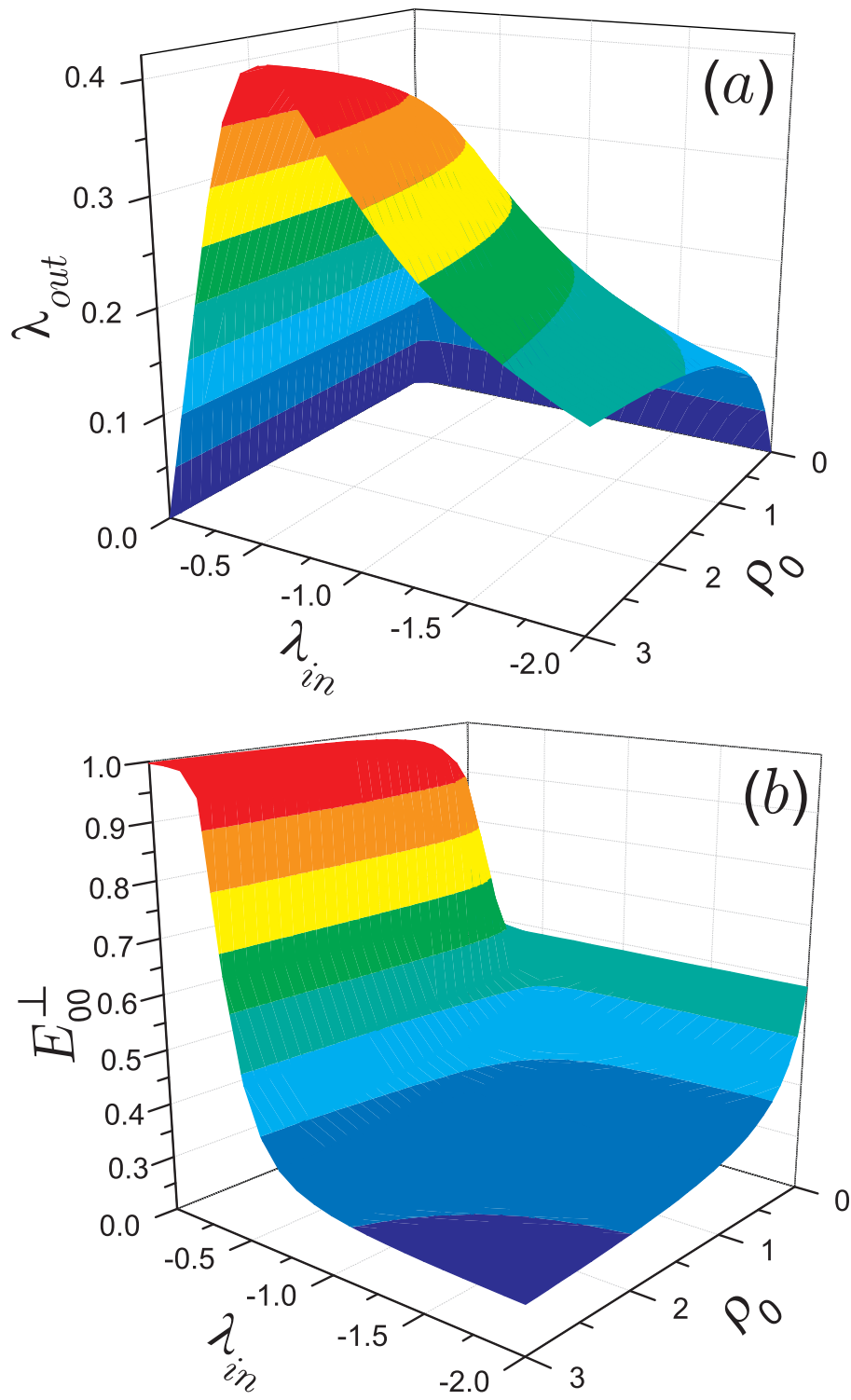}
\caption{\label{Fig6}
(a)  One of the surfaces in $(\rho_0,\lambda_{in},\lambda_{out})$ space on which the energy $E_{00}^\perp$ is real. (b) {\em Real} energy $E_{00}^\perp$ corresponding to panel (a) as a function of $\rho_0$ and $\lambda_{in}$.}
\end{figure}

Next, let us address the important question about the conditions under which the 2D energy $E_{\Phi;nm}^\perp$ is real. Apart from the theoretical interest related, for example, to the actively developing field of $\cal PT$-symmetric quantum mechanics \cite{Bender1,Bender2,Bender3}, this problem is of a paramount practical applications since real energy means a lossless longitudinal current down the waveguide. As it follows from Eq.\ (\ref{BesselEq1}), the energy $E_{\Phi;nm}^\perp$, in order to be {\em real}, has to satisfy at the same time the following equations:
\begin{subequations}\label{RealEnergy1}
\begin{eqnarray}\label{RealEnergy1A}
\lambda_{in}{\cal F}_{m_\Phi}^{ND}\left(\pi\sqrt{E_{\Phi;nm}^\perp}\right)-\lambda_{out}{\cal F}_{m_\Phi}^{DN}\left(\pi\sqrt{E_{\Phi;nm}^\perp}\right)=0,\\
\label{RealEnergy1B}
\lambda_{in}\lambda_{out}\pi^2E_{\Phi;nm}^\perp{\cal F}_{m_\Phi}^{NN}\left(\pi\sqrt{E_{\Phi;nm}^\perp}\right)-{\cal F}_{m_\phi}^{DD}\left(\pi\sqrt{E_{\Phi;nm}^\perp}\right)=0,
\end{eqnarray}
\end{subequations}
defining different branches of its $\lambda_{in}$-, $\lambda_{out}$- and $\rho_0$-dependence. Since for the fixed de Gennes distances and inner radius the {\em real} energy $E_{\Phi;nm}^\perp$ has to obey {\em two} equations (\ref{RealEnergy1}) {\em simultaneously}, this happens only for some definite correlations between $\lambda_{in}$, $\lambda_{out}$ and $\rho_0$.
For example, it is immediately seen from Eqs.~(\ref{Asymptotics1}) that in the asymptotic cases the energy is real when the following conditions hold:
\begin{subequations}\label{Asymptotics3}
\begin{eqnarray}\label{Asymptotics3DD}
\lambda_{in}{\cal F}_{m_\Phi}^{ND}\left(x_{m_\Phi n}^{DD}\right)=\lambda_{out}{\cal F}_{m_\Phi}^{DN}\left(x_{m_\Phi n}^{DD}\right),\quad\left|\lambda_{in,out}\right|\ll 1,\qquad\qquad\\
\label{Asymptotics3DN}
\left(x_{m_\Phi n}^{DN}\right)^2\lambda_{in}{\cal F}_{m_\Phi}^{NN}\left(x_{m_\Phi n}^{DN}\right)=-\left.{\cal F}_{m_\Phi}^{DD}\left(x_{m_\Phi n}^{DN}\right)\right/\lambda_{out},\quad\left|\lambda_{in}\right|\ll 1,\left|\lambda_{out}\right|\gg 1,\qquad\qquad\\
\label{Asymptotics3ND}
\left.{\cal F}_{m_\Phi}^{DD}\left(x_{m_\Phi n}^{ND}\right)\right/\lambda_{in}=\left(x_{m_\Phi n}^{ND}\right)^2\lambda_{out}{\cal F}_{m_\Phi}^{NN}\left(x_{m_\Phi n}^{ND}\right),\quad\left|\lambda_{in}\right|\gg 1,\left|\lambda_{out}\right|\ll 1,\qquad\qquad\\
\label{Asymptotics3NN}
\left.\begin{array}{cl}\left.{\cal F}_{m_\Phi}^{DN}\left(x_{m_\Phi n}^{NN}\right)\right/\lambda_{in}=\left.{\cal F}_{m_\Phi}^{ND}\left(x_{m_\Phi n}^{NN}\right)\right/\lambda_{out},&\left|m_\Phi\right|+n\ne 0,\\ \rho_0\left/\lambda_{in}\right.=-\rho_1\left/\lambda_{out}\right.,& m_\Phi=n=0,\end{array}\right\}\left|\lambda_{in,out}\right|\gg 1.\qquad\qquad
\end{eqnarray}
\end{subequations}
In turn, in the limit of the large radius, $\rho_0\rightarrow\infty$, Eqs. (\ref{RealEnergy1}) transform to
\begin{subequations}\label{RealEnergy2}
\begin{eqnarray}\label{RealEnergy2A}
\left(\lambda_{in}+\lambda_{out}\right)\cos\pi\sqrt{E_n^\perp}&=&0,\\
\label{RealEnergy2B}
\left(\lambda_{in}\lambda_{out}\pi^2E_n^\perp+1\right)\sin\pi\sqrt{E_n^\perp}&=&0.
\end{eqnarray}
\end{subequations}
They also follow straightforwardly from the eigenvalue equation of the straight waveguide with Robin boundary conditions, Eq.~(\ref{StraightThresholds1}). Then, one has following situations
\begin{subequations}\label{RealEnergy3}
\begin{eqnarray}\label{RealEnergy3A}
E^\perp= \frac{1}{\pi^2\lambda_{in}^2},\ n^2, n=1,2,\ldots\quad {\rm for}\quad\lambda_{out}=-\lambda_{in},\\
\label{RealEnergy3B}
E_n^\perp=(n+1/2)^2,n=0,1,\ldots\quad{\rm for}\quad\lambda_{in}\lambda_{out}=-\frac{1}{\pi^2\left(n+1/2\right)^2}.
\end{eqnarray}
\end{subequations}
Spectrum from Eq.\ (\ref{RealEnergy3A}) has been known before \cite{Krejcirik1,Krejcirik2,Krejcirik3,Borisov1,Krejcirik4} from $\cal PT$-symmetric \cite{Bender1,Bender2,Bender3} study of the quantum systems. This allows to understand the features of one of the sheets of the real energy surface in $(\rho_0,\lambda_{in},\lambda_{out})$ coordinates shown in Fig. \ref{Fig6}(a). Namely, at the large radius the ground state has, for the small $|\lambda_{in}|$, as it follows from (\ref{RealEnergy3A}), the real energy $E^\perp=1$ under the condition $\lambda_{out}=-\lambda_{in}$. As the inner Robin parameter passes the value of $|\lambda_{in}|=1/\pi$, the ground energy becomes $E_0^\perp=1/\left(\pi\lambda_{in}\right)^2$, and the same condition for the inner and outer Robin distances holds. This steep energy descent is clearly seen in panel (b) of Fig. \ref{Fig6} exhibiting corresponding {\em real} energies $E_{00}^\perp$. As the inner de Gennes distance goes through $|\lambda_{in}|=2/\pi$, the ground energy becomes $E_0^\perp=1/4$, and the relation between $\lambda_{in}$ and $\lambda_{out}$ reads: $\lambda_{out}=-4/\left(\pi^2\lambda_{in}\right)$. In other words, the described evolution is a transformation from the purely Dirichlet boundary condition $\lambda_{in}=\lambda_{out}=0$ with the energy spectrum being squares of integers to its Dirichlet-Neumann counterpart $1/\lambda_{in}=\lambda_{out}=0$ with the spectrum of squares of half-integers through such intermediate values of $\lambda_{in}$ and $\lambda_{out}$ which guarantee a zero value of the imaginary part of the energy. Thus, the extra (compared to the pure Dirichlet case) state with the energy $1/(\pi\lambda)^2$ in Eq.\ (\ref{RealEnergy3A}) plays the role of the bridge linking two real-energy levels with the different types of the requirements at the walls. As Eqs.\ (\ref{RealEnergy1}) and (\ref{Asymptotics3}) demonstrate and Fig.\ \ref{Fig6} depicts, the similar dependence persists for the finite $\rho_0$ too, however, the coefficients between the inner and outer purely imaginary de Gennes distances are now radius dependent. Of course, for the solid cylinder, $\rho_0=0$, the real energies $E_{\Phi;nm}^\perp=(j_{|m_\Phi|n}/\pi)^2$ are possible only for $\lambda_{out}\equiv 0$.
\begin{figure}
\centering
\includegraphics[width=0.99\columnwidth]{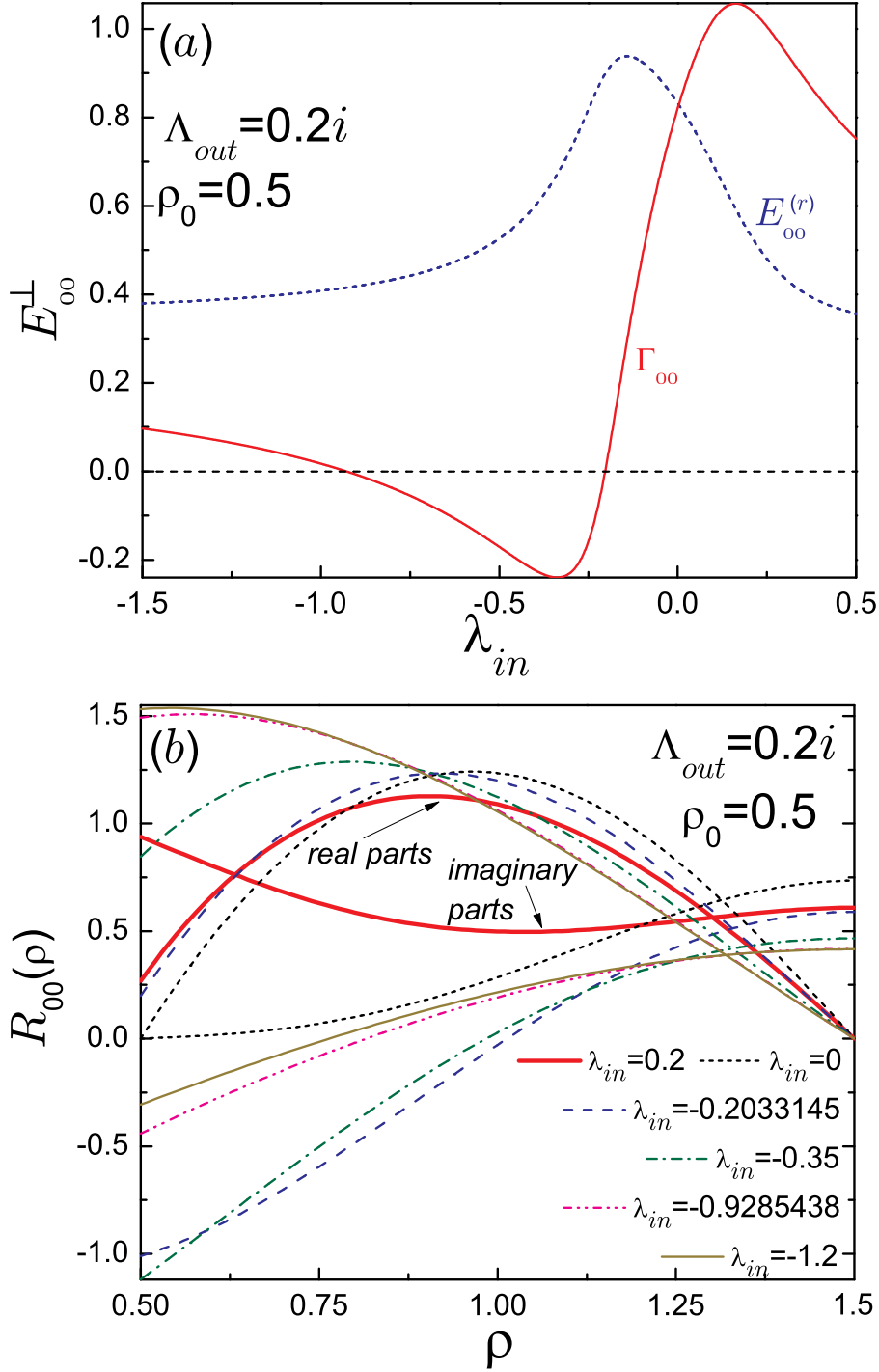}
\caption{\label{Fig7}
(a) Complex energy $E_{00}^\perp$ as a function of the inner purely imaginary de Gennes distance $\lambda_{in}$ for $\rho_0=0.5$ and $\lambda_{out}=0.2$ where the dotted line depicts the real part of the energy $E_{00}^{(r)}$ while the solid line is for the double negative imaginary part. Dashed horizontal line denotes zero energy. (b) Real (lower at $\rho\approx\rho_0+1$) and imaginary (upper at $\rho\approx\rho_0+1$) parts of the total radial functions $R_{00}(\rho)$ for $\rho_0=0.5$, $\lambda_{out}=0.2$ and several $\lambda_{in}$  where the thick solid lines are for $\lambda_{in}=0.2$, dotted lines - for $\lambda_{in}=0$, dashed lines  - for $\lambda_{in}=-0.2033145$, dash-dotted lines  - for $\lambda_{in}=-0.35$, dash-dot-dotted lines  - for $\lambda_{in}=-0.9285438$, and thin solid lines are for  $\lambda_{in}=-1.2$. Note that real parts vanish identically at $\rho=\rho_0+1$.}
\end{figure}

To further elaborate on this issue, in Fig. \ref{Fig7} the complex energy $E_{00}^\perp$ [panel (a)] and the corresponding radial function $R_{00}(\rho)$ [panel (b)] are shown as functions of the inner de Gennes distance $\lambda_{in}$ for the fixed outer extrapolation length $\lambda_{out}$ and radius $\rho_0$. For the given fixed parameters, the energy turns real at the two imaginary Robin lengths: $\lambda_{in}=-0.9285438$ and $\lambda_{in}=-0.2033145$. The form of the real parts of the corresponding radial functions in panel (b) manifests that first of them is close to the Neumann-Dirichlet edge of the spectrum while the main contribution to the second one is from the pure Dirichlet configuration, as it was discussed in the previous paragraph. To turn $\Gamma$ into zero, each of the functions is deformed compared to the corresponding type of the boundary requirements and, in addition, contains an admixture of the imaginary component. However, even though the energies for these two parameters are real, the transversal radial current through the ring exists. This directly follows from the form of the functions in panel (b) of Fig.\ \ref{Fig7} and expression (\ref{current3}) for the total radial current which for our parameters reads:
\begin{equation}\label{current4}
J_{\rho_{0,1}}=\pm\frac{\rho_{0,1}}{\lambda_{in,out}}\left|R\left(\rho_{0,1}\right)\right|^2
\end{equation}
with plus (minus) sign corresponding to the inner (outer) surface. Note that in general case of the complex $\Lambda$, the expressions for these currents do not depend on the sign of the real parts of the extrapolation lengths:
\begin{equation}\label{current5}
J_{\rho_{0,1}}=\pm\rho_{0,1}\left|R\left(\rho_{0,1}\right)\right|^2\frac{{\rm Im}\left(\Lambda_{in,out}\right)}{\left|\Lambda_{in,out}\right|^2}.
\end{equation}
The crucial point here is the fact that the equality of these two fluxes forces the energy to be real:
\begin{equation}\label{conservation2}
\left.\Gamma\right|_{J_{\rho_0}=J_{\rho_1}}\equiv 0.
\end{equation}
In fact, Eq.\ (\ref{conservation2}) is a reflection of the current conservation law when the incoming flux through one of the surfaces is equal to the outgoing one and, thus, the longitudinal current is not affected by the processes occurring in the cross section of the waveguide. Expression for the divergence, Eq.\ (\ref{divergence1}), tells that every spatial point, including that at the annulus inner and outer circumferences, absorbs and emits the same amount of the flux and, thus, the total flow through it is zero. The difference with the systems with real Robin parameters is in the fact that in the latter situation {\em no} current passes at all through the points lying on the boundary. It immediately follows from equations (\ref{current5}) and (\ref{conservation2}) that for the real energy the following condition holds:
\begin{equation}\label{ZeroGamma1}
\left.\left[{\rm Im}\left(\Lambda_{in}\right){\rm Im}\left(\Lambda_{out}\right)\right]\right|_{\Gamma=0}<0.
\end{equation}
\begin{figure}
\centering
\includegraphics[width=0.96\columnwidth]{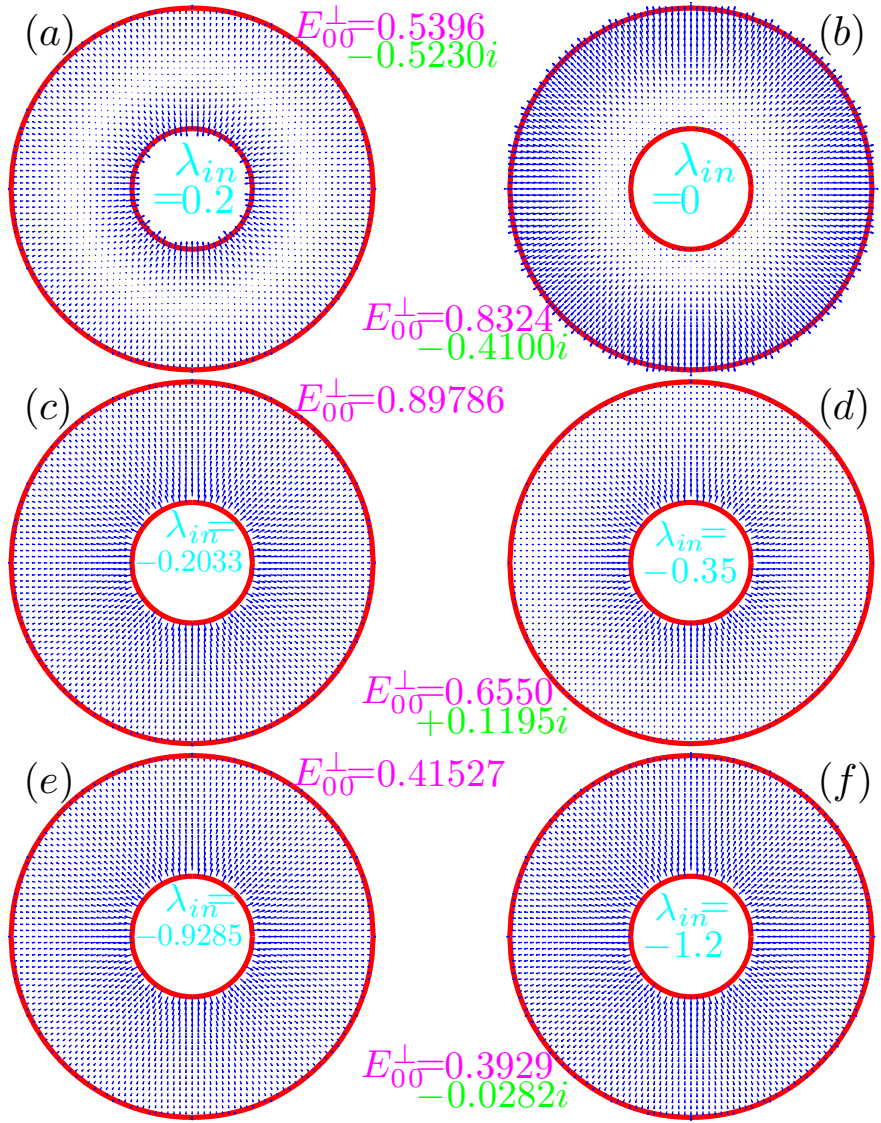}
\caption{\label{Fig8}
Current densities of the state with $n=m=\Phi=0$ for $\rho_0=0.5$, $\Lambda_{out}=0.2i$ and (a) $\Lambda_{in}=0.2i$, (b) $\Lambda_{in}=0$, (c) $\Lambda_{in}=-0.2033145i$, (d) $\Lambda_{in}=-0.35i$, (e) $\Lambda_{in}=-0.9285438i$, and  (f) $\Lambda_{in}=-1.2i$. For each of the figures, the currents are normalized with respect to their largest value. The longer arrows denote larger currents. Corresponding energies $E_{00}^\perp$ are written near each pattern.}
\end{figure}

Current density patterns corresponding to the wave functions from Fig.\ \ref{Fig7}(b) are shown in Fig.\ \ref{Fig8}. It is seen from panel (a) that for the same imaginary extrapolation lengths (or, more generally, for the de Gennes distances with the same signs of their imaginary parts, $\lambda_{in}\lambda_{out}>0$) the currents through the inner and outer interfaces flow in the opposite radial directions; accordingly, the annulus acts as a sink (source) for positive (negative) $\lambda$ with nonzero divergence of the current, and, as it follows from Eq.\ (\ref{divergence1}), no real energies can be obtained. The same is true for the configuration when one of the surfaces supports real Robin requirement. For example, even though the divergence at the Dirichlet inner border in panel (b) is zero due to the vanishing function on it, the outer reactively scattering wall does not allow to suppress the imaginary part of the transverse energy $E_{00}^\perp$. In turn, for the real energies [panels (c) and (e)] the fluxes through the inner and outer circle head in the same direction and, since they exactly compensate each other, $J_{\rho_0}=J_{\rho_1}$, one gets a lossless longitudinal current with zero divergence of its transverse counterpart. Any deviation from these two points results in unequal currents through the confining surfaces when, depending on the ratio between imaginary de Gennes lengths with the opposite signs, the ring acts as a sink [panel (d)] or a source [panel (f)].

\subsubsection{Interplay of the magnetic field and imaginary de Gennes distance}
\begin{figure}
\centering
\includegraphics[width=0.96\columnwidth]{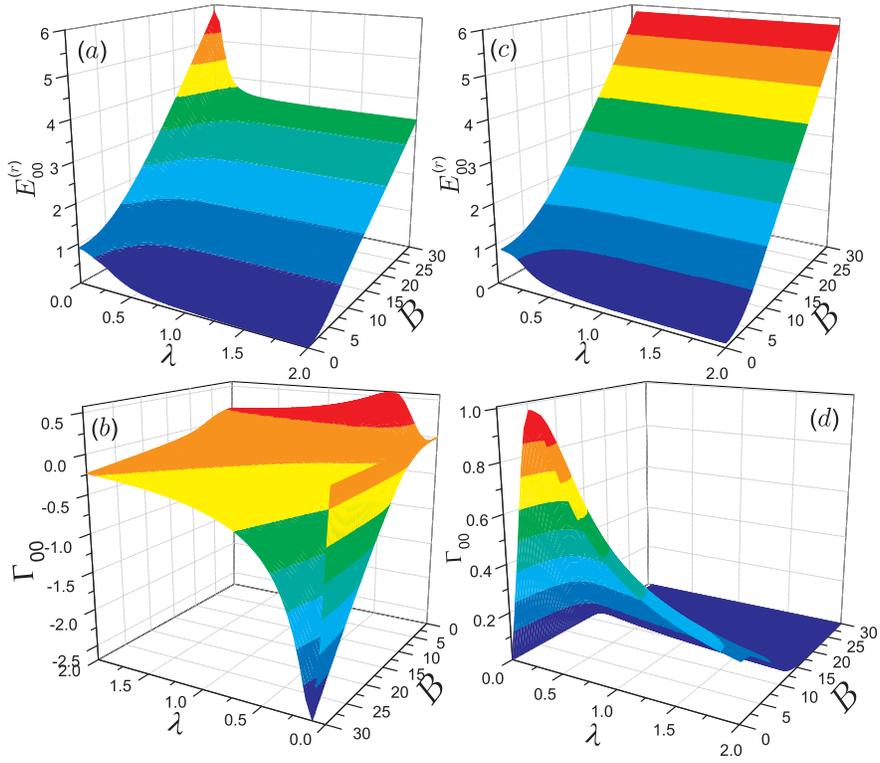}
\caption{\label{Fig9}
Real [panels (a) and (c)] and double negative imaginary [panels (b) and (d)] parts of the total transverse energy $E_{00}^\perp$ as functions of the purely imaginary Robin parameter $\lambda$ and field $B$ for the annulus with inner radius $\rho_0=0.1$. For the left panels $\lambda=\lambda_{out}=-\lambda_{in}$ while for the right panels $\lambda=\lambda_{out}$ with $\lambda_{in}=0$. Note reversed $\lambda$ and $B$ axes in panel (b) as compared to other panels.}
\end{figure}

For the quantum dot, the increasing magnetic field pushes the charged carrier to the centre of the disk, i.e., away from the Robin boundary with the corresponding decrease of its impact on the properties of the system; as a result, the resonance on the $\Gamma-{\rm Im}(\Lambda)$ characteristics   gets smaller with growing $B$ and, ultimately, in the limit of the very strong intensities, one recovers the states with independent of $\Lambda$ real energies in the form of the Landau levels, Eq.\ (\ref{LandauLevels1}) \cite{Olendski1}. For the ring geometry with nonzero $\rho_0$ the formation of the Landau states is prevented by the inner impenetrable wall which, in the case of the growing fields, plays the more and more dominant role since the distribution of the parameter $\Psi({\bf r})$ concentrates around it. Accordingly, one can expect the enhancement of the reactive scattering for the nonzero value of the imaginary part of its Robin parameter. To exemplify this, we present in Fig.\ \ref{Fig9} the energies $E_{00}^\perp$ of the annulus with radius $\rho_0=0.1$ for the two configurations of the  extrapolation length $\lambda$: the left panels show real and imaginary parts of the total transverse energy for the opposite signs of the Robin parameters on the inner and outer surfaces, $\lambda=\lambda_{out}=-\lambda_{in}$, while for the right panels the Dirichlet boundary condition was adopted on the inner wall: $\lambda\equiv\lambda_{out}$, $\Lambda_{in}=0$. At the zero field, the real parts $E_{\Phi;nm}^{(r)}$ of the energy  show with the increasing magnitude of the de Gennes distance a transformation from the Dirichlet case at $\lambda=0$ with $E_{\Phi;nm}^\perp=\left(x_{\left|m_\Phi\right|n}^{DD}/\pi\right)^2$ to the either pure Neumann [panel (a)], $E_{\Phi;nm}^\perp=\left(x_{\left|m_\Phi\right|n}^{NN}/\pi\right)^2$, or Dirichlet-Neumann [panel (c)], $E_{\Phi;nm}^\perp=\left(x_{\left|m_\Phi\right|n}^{DN}/\pi\right)^2$, geometry at the large Robin parameter, $\lambda\gg 1$. Imaginary parts at $B=0$ exhibit discussed above variation of $\Gamma_{\Phi;nm}$ with its passage through zero for the opposite signs of $\lambda$ [panel (b)] or the resonant dependence caused by the outer imaginary extrapolation length [panel (d)]. Growing magnetic field restricts the motion to the area around the inner surface with the corresponding change of the energies $E_{\Phi;nm}^\perp$; for example, for the inner negative $\lambda$ one sees in panels (a) and (c) of Fig.\ \ref{Fig9} that, for the strong fields, the real part $E_{\Phi;nm}^{(r)}$ varies with the de Gennes distance only for $\lambda\lesssim 2/B$, while the $\Gamma$ dependence transforms into the resonance profile with almost quadratic dependence on $B$ - the behaviors familiar from the study of the columnar defect, section \ref{Section3_1_2}. For the Dirichlet inner wall in the increasing field, the particle moves away from the outer scattering Robin interface with the corresponding decrease of its influence on the system; accordingly, the $\Gamma$ dependence on $\lambda$ quenches and, at the large $B$, tends to zero for all external extrapolation lengths [panel (d)]. At the same time, the real part grows with the field with its dependence on $\lambda$ fading too, and in the limit of the strong fields such that $B\rho_0^2\gg 1$, one gets:
\begin{equation}\label{PillarEq6}
\left.E_{\Phi;nm}^\perp\right|_{\Lambda_{in}=0}=\frac{1}{4\pi^2}\left\{1+\left[\left(n+\frac{3}{4}\right)\frac{6\pi}{B\rho_0^2}\right]^{2/3}\right\}\left(B\rho_0\right)^2+\frac{m_\Phi}{\pi^2}B, \quad B\rho_0^2\rightarrow\infty
\end{equation}
(recall that we use different length units here and in subsection \ref{ColumnarDefect1}).
In our model of the electric field influence on superconductors, Eqs.\ (\ref{TotalExtrapolationLength1}) and (\ref{PotentialU}),  it means that the combined application of the external electric $\mbox{\boldmath$\cal E$}$ and magnetic $\bf B$ fields allows to vary the scattering in a broad range switching from the lossless current to any desired degree of scattering by the simple change of the gate voltage and/or magnetic intensity. Comparing inner Dirichlet case of the ring with the energy $E_{\phi;nm}^\perp$ dependence on ${\rm Im}(\Lambda)$ and $B$ for the disk (Fig. 2 in Ref.\ \cite{Olendski1}), one sees their qualitative and, for the imaginary part, almost quantitative, similarity with the difference being in the $E_{\Phi;nm}^{(r)}$ dependence on the field for the large $B$: linear, Eq.\ (\ref{LandauLevels1}), for the disk and quadratic, Eq.\ (\ref{PillarEq6}), for the annulus.

\section{Concluding remarks}
\label{sec4}
Ubiquitousness in Nature of the physical systems which are described by the wave equation (\ref{WaveEquation1}) supplemented by the Robin boundary condition, Eq.\ (\ref{Robin1}), with complex extrapolation length $\Lambda$ dictates a necessity of their correct theoretical description. In the present research, the previous results for the simply connected domain of the 2D circular disk and 3D solid cylinder \cite{Olendski1} were supplemented by the calculations of the simplest doubly connected geometry of the 2D ring and infinitely long 3D cylinder of the annular cross section with, in general, different complex Robin lengths on the two confining circumferences. Comparative analysis between the two configurations revealed new features for the multiply connected structures; for example, it was predicted that under some correlation between the outer and inner Robin distances with the opposite signs of their imaginary parts, the eigenvalue of Eq.\ (\ref{WaveEquation1}) turns real - despite of the complexness of the boundary condition (\ref{Robin1}). Physically, this is explained by the same amount of the flux entering the system through one interface and leaking out from it through the other one. This result, interesting by itself for, e.g., the $\cal PT$-symmetric quantum mechanics \cite{Bender1,Bender2,Bender3}, is of a large practical significance too since the {\em real}  transverse energy $E_{\Phi;nm}^\perp$ restores lossless longitudinal current down the waveguide. New phenomena are predicted to exist for the system response to the external magnetic field $\bf B$ too; namely, if, the growing $B$ for the disk decreases the imaginary part of the complex transverse energy turning, for the large intensities, the levels into the Landau states, then, for the 2D ring, $\Gamma$ fades with the increasing field only if the inner Robin parameter is real, otherwise, due to the reactive scattering at the inner wall, the imaginary part of the energy grows quadratically with $B$.

We considered the hollow structures  invariant under  the rotation around the $z$ axis. In the process of the sample growth intentional or unintentional deviations from the azimuthal symmetry can occur. Such Neumann structures have been studied both theoretically \cite{Berger1,Berger2,Vodolazov1} and experimentally \cite{Morelle1,Furugen1}. The same Dirichlet geometry was used for the study of the distorted semiconductor rings \cite{BrunoAlfonso1,BrunoAlfonso2}. Combined influence of the excentricity, real de Gennes distance and uniform magnetic field $\bf B$ was also calculated \cite{Moncada1}. Based on the results presented above, the conjecture can be made that if the displacement $\rho_{sh}$ of the inner disk from the centre of the outer one is larger than its radius $\rho_0$, then for the high enough fields when the ratio $r_B/(\rho_{sh}-\rho_0)$ is smaller than unity, the imaginary part of the energy will asymptotically tend to zero with its real part approaching the Landau levels, Eq.\ (\ref{LandauLevels1}). In the opposite case, $\rho_{sh}<\rho_0$, the $\Gamma-B$ characteristics at the strong fields will depend on the imaginary part of the inner extrapolation length $\Lambda_{in}$; e.g., it will decrease to zero for $B\rightarrow\infty$ at ${\rm Im}(\Lambda_{in})=0$ and its magnitude will almost quadratically increase for any nonzero ${\rm Im}(\Lambda_{in})$ while the real part of the energy $E^{(r)}$ in either case will exhibit the same $\sim B^2$ dependence. The waveguides with more complicated cross sections (with, for example, two or more leaking holes inside the larger circle with complex Robin parameter) can be treated in a way being a combination of the methods developed here and in the cited references, and, depending on the particular geometry, they should exhibit one of the dependencies just described.

\bibliographystyle{model1a-num-names}

\end{document}